\documentclass[12pt]{article}
\usepackage{a4wide}
\usepackage{amsmath,amsthm,amsfonts}
\usepackage{graphicx}

\newtheorem{theorem}{Theorem}[section]

\newtheorem{remark}[theorem]{Remark}

\newcommand{\url}[1]{{\tt \small #1}}
\newcommand{\Ex}[2]{\mathbb{E}_{#1}\!\left[\,#2\,\right]}
\newcommand{\Qx}[1]{\mathbb{Q}\left\{\,#1\,\right\}}

\newcommand{\defleg}{\mbox{D{\tiny EF}L{\tiny EG}}}
\newcommand{\prmleg}{\mbox{P{\tiny REMIUM}L{\tiny EG}}}

\newcommand{\gpl}{Z}

\newcommand{\defrate}{C}
\newcommand{\defratenorm}{{\bar\defrate}}
\newcommand{\loss}{L}
\newcommand{\lossnorm}{{\bar\loss}}
\newcommand{\losstr}{\lossnorm^{A,B}}
\newcommand{\cds}{S}
\newcommand{\cdstr}{S^{A,B}}
\newcommand{\upfronttr}{U^{A,B}}
\newcommand{\ind}[1]{1_{\{#1\}}}
\newcommand{\rec}{R}

\title{{\large \bf Default correlation, cluster dynamics and single names: \\ The GPCL dynamical loss model}
\thanks{This paper has been written partly as a response to criticism, suggestions, encouragements and objections to our earlier GPL paper.
In particular, we are grateful to Aur\'{e}lien Alfonsi, Marco Avellaneda, Norddine Bennani, Tomasz Bielecki, Giuseppe Castellacci, Dariusz Gatarek, Diego Di Grado, Youssef Elouerkhaoui, Kay Giesecke, Massimo Morini, Chris Rogers and Lutz Schl{\"o}gl for helpful comments, criticism, further references and suggestions.} 
}
\author{Damiano Brigo \ \ \ Andrea Pallavicini \ \ \  Roberto Torresetti \\
        Credit Models - Banca IMI.
        Corso Matteotti 6 - 20121 Milano, Italy \\
\url{\{damiano.brigo,andrea.pallavicini,roberto.torresetti\}@bancaimi.it} \\ \\
{\small Available at  {\tt http://www.damianobrigo.it} }\\
{\normalsize Updated version in the International Journal of Theoretical and Applied Finance}\\{\normalsize and in the book "Credit Correlation: Life after Copulas" by World Scientific, 2007} }
\date{\small First Version: November 20, 2006. First Posted on Jan 12, 2007 at SSRN. 
}
\begin{document}

\maketitle

\begin{abstract}

We extend the common Poisson shock framework reviewed for example
in Lindskog and McNeil (2003) to a formulation avoiding repeated
defaults, thus obtaining a model that can account consistently for
single name default dynamics, cluster default dynamics and default
counting process. This approach allows one to introduce
significant dynamics, improving on the standard ``bottom-up"
approaches, and to achieve true consistency with single names,
improving on most ``top-down" loss models. Furthermore, the
resulting GPCL model has important links with the previous GPL
dynamical loss model in Brigo, Pallavicini and Torresetti
(2006a,b), which we point out. Model extensions allowing for more
articulated spread and recovery dynamics are hinted at.
Calibration to both DJi-TRAXX and CDX index and tranche data across
attachments and maturities shows that the GPCL model has the same
calibration power as the GPL model while allowing for consistency
with single names.
\end{abstract}

{\bf JEL classification code: G13. \\ \indent AMS classification codes: 60J75, 91B70}

\medskip


\newpage
{\small \tableofcontents}
\newpage

\section{Introduction}

The modeling of dependence or ``correlation" between the default
times of a pool of names is the key issue in pricing financial
products depending in a non-linear way on the pool loss.
Typical examples are CDO tranches, forward start CDO's and tranche
options.

\subsubsection*{Bottom-up approach}

A common way to introduce dependence in credit derivatives
modeling, among other areas, is by means of copula functions. A
copula corresponding to some preferred multivariate distribution
is ``pasted" on the exponential random variables triggering
defaults of the pool names according to first jumps of Poisson or
Cox processes. In general, if one tries to control dependence by
specifying dependence across single default times, one is
resorting to the so called ``bottom-up" approach, and the copula
approach is typically within this framework. Yet, such procedure
cannot be extended in a simple way to a fully dynamical model in
general. A direct alternative is to insert dependence among the
default intensities dynamics of single names either by direct
coupling between intensity processes or by introducing common
factor dynamics. See for example the paper by Chapovsky, Rennie
and Tavares (2006).

\subsubsection*{Top-down approach}

On the other side, one could give up completely single default
modeling and focus on the pool loss and default counting
processes, thus considering a dynamical model at the aggregate
loss level, associated to the loss itself or to some suitably
defined loss rates. This is the ``top-down" approach pioneered by
Bennani (2005, 2006), Giesecke and Goldberg (2005),
Sidenius, Piterbarg and Andersen (2005), Sch\"onbucher (2005), Di
Graziano and Rogers (2005), Brigo, Pallavicini and Torresetti
(2006a,b), Errais, Giesecke and Goldberg (2006) among others.
The first joint calibration results of a single model
across indices, tranches attachments and maturities, available in
Brigo, Pallavicini and Torresetti (2006a), show that even a
relatively simple loss dynamics, like a capped generalized Poisson
process, suffices to account for the loss distribution dynamical
features embedded in market quotes. However, to justify the
``down" in ``top-down" one needs to show that from the aggregate
loss model, possibly calibrated to index and tranche data, one can
recover a posteriori consistency with single-name default
processes. Errais, Giesecke and Goldberg (2006) advocate the use
of random thinning techniques for their approach, but in general
it is not clear whether a fully consistent single-name default
formulation is possible given an aggregate model as the starting point.
Interesting research on this issue is for example in Bielecki, Vidozzi and Vidozzi (2007), who play on markovianity of families of single name and multi-name processes with respect to different filtrations, introducing assumptions that limit the model complexity needed to ensure consistency.

\subsubsection*{Common Poisson Shock (CPS) approach and Marshall-Olkin copula}

Apart from these two general branches and their problems, mostly
the above mentioned lack of dynamics in the classical ``bottom-up"
approach and the possible lack of ``down" in the ``top-down"
approach, there is a special ``bottom-up" approach that can lead
to a loss dynamics resembling some of the ``top-down" approaches
above, and the model in Brigo, Pallavicini and Torresetti (2006a)
in particular. This approach is based on the common Poisson shock
(CPS) framework, reviewed in Lindskog and McNeil (2003) with
application in operational risk and credit risk for very large
portfolios. This approach allows for more than one defaulting name
in small time intervals, contrary to some of the above-mentioned
``top-down" approaches.

The problem of the CPS framework is that it leads in general to
repeated defaults. If one is willing to assume that single names
and groups of names may default more than once, actually infinite
times, the CPS framework allows one to model consistently single
defaults and clusters defaults. Indeed, if we term ``cluster" any
(finite) subset of the (finite) pool of names, in the CPS
framework different cluster defaults are controlled by independent
Poisson processes. Starting from the clusters defaults one can
easily go back either to single name defaults (``bottom-up") or to
the default counting process (``top-down"). Thus we have a consistent
framework for default counting processes and single name default,
driven by independent clusters-default Poisson processes. In the
``bottom-up" language, one sees that this approach leads to a
Marshall-Olkin copula linking the first jump (default) times of
single names.  In the ``top-down" language, this model looks very
similar to the GPL model in Brigo, Pallavicini and Torresetti
(2006a) when one does not limit the number of defaults.

In the credit derivatives literature the CPS framework has been
used for example in Elouerkhaoui (2006), see also references
therein. Balakrishna (2006) introduces a semi-analytical approach
allowing again for more than one default in small time intervals
and hints at its relationship with the CPS framework, showing also
some interesting calibration results.

\subsubsection*{CPS without repeated defaults?}

Troubles surface when one tries to get rid of the unrealistic
``repeated default" feature. In past works it was argued that one
just assumes cluster default intensities to be small, so that the
probability that the Poisson process for one cluster jumps more
than once is small. However, calibration results in Brigo,
Pallavicini and Torresetti (2006a) lead to high enough intensities
that make repeated defaults troublesome. The issue remains then if
one is willing to use the CPS framework for dependence modeling in
credit derivatives pricing and hedging.

\subsubsection*{New results}

In this paper we start from the standard CPS framework with
repeated defaults and use it as an engine to build a new model for
(correlated) single name defaults, clusters defaults and default
counting process or portfolio loss. Indeed, if $s$ is a set of names
in the portfolio and $|s|$ is the number of names in the set $s$,
we start from the (independent) cluster default Poisson processes
$\widetilde{N}_s$ for example in Lindskog and McNeil (2003),
consistent with (correlated) single name $k$ repeated default
Poisson processes $N_k$, and build new default processes
avoiding repeated single name and cluster defaults. We propose two
ways to do this, the most interesting one leading to a new
definition of cluster defaults ${\widetilde{N}^2}_s$ avoiding
repetition and (correlated) single name defaults $N^2_k$
avoiding repetition as well, whose construction is detailed in
Section~\ref{sec:consistent-selfaffecting}. An alternative
approach, based on an adjustment to avoid repeated defaults at
single name level, and leading to (correlated) single name default
processes $N^1_k$, is proposed in
Section~\ref{sec:ncheck-notcons}. This approach however leads to a
less clear cluster dynamics in terms of the original cluster
repeated default processes $\widetilde{N}_s$.

\subsubsection*{The Generalized Poisson Cluster Loss (GPCL) model}

We then move on and examine the approach based on the non-repeated
cluster and single name default processes
${\widetilde{N}^2}_s,N^2_k$, which we term ``Generalized Poisson
Cluster Loss model" (GPCL), detailing some homogeneity assumptions
that can reduce the otherwise huge number of parameters in this
approach. We calibrate the associated default counting process to
a panel of index and tranche data across maturities, and compare
the resulting model with the Generalized Poisson Loss (GPL) model
in Brigo, Pallavicini and Torresetti (2006a,b). The GPL model
is similar to the GPCL model but lacks a clear interpretation in
``bottom-up" terms, since we act on the default counting process, by
capping it to the portfolio size, without any control of what
happens either at single name or at clusters level. The GPCL
instead allows us to understand what happens there. Calibration
results are similar but now we may interpret the ``top-down" loss
dynamics associated to the default counting process in a ``bottom-up"
framework, however stylized this is, and have a clear
interpretation of the process intensities also in terms of default
clusters.

\subsubsection*{Possible extensions}

In Section~\ref{sec:extensionsgpcl} we present possible extensions
leading to richer spread dynamics and recovery specifications.
This, in principle, allows for more realism in valuation of
products that depend strongly on the spread dynamics such as
forward starting CDO tranches or tranche options.  However, since
we lack liquid market data for these products, we cannot proceed
with a thorough analysis of the extensions. Indeed, the extensions
are only presented as a proposal and to illustrate the fact that
the model is easily generalizable. Further work is in order when
data will become available.

\section{Modeling framework and the Common Poisson Shock approach}

We consider a portfolio of $M$ names, typically $M=125$, each with
notional $1/M$ so that the total pool has unit notional. We denote
with $\lossnorm_t$ the portfolio cumulated loss, with $\defrate_t$
the number of defaulted names up to time $t$ (``default counting
process") and we define $\defratenorm_t := \defrate_t/M$ (default
rate of the portfolio).

Since a portion of the amount lost due to each default is usually
recovered, the loss is smaller than the default fraction. Thus,
\begin{equation}\label{eq:no-arb-const}
 0 \le d\lossnorm_t\le d  \defratenorm_t\le 1 \ \ \ \mbox{for all} \ \ \
 t,\ \ \ \lossnorm_0 = 0, \defratenorm_0 = 0,
\end{equation}
which in turn implies (but is {\em not} implied by) $0 \le
\lossnorm_t\le \defratenorm_t\le 1$.

Notice that with the notation $d X_t$, where $X_t$ is a jump process which we assume to be right continuous with left limit, we actually mean the jump size of process $X$ at time $t$ if $X$ jumps at $t$, and zero otherwise, or, in other terms, pathwise, $d X_t = X_t - X_{t^-}$, where in general we define $X_{t^-} :=\lim_{h \downarrow 0} X_{t-h}$.

We can relate the cumulated
loss process $\lossnorm_t$ and the re-scaled number of defaults
$\defratenorm_t$ at any time $t$ through the notion of {\it
recovery rate at default} $\rec_t$,
\begin{equation}
\label{eq:afc} d\lossnorm_t = d\defratenorm_t (1-\rec_t)
\end{equation}
where $\rec_t$ satisfies some technicalities that we detail later
in Section~\ref{sec:noabrconst-rec}. This equation actually is an abbreviation for
\[  \lossnorm_t = \int_0^t   (1-\rec_u) d\defratenorm_u . \]

The no-arbitrage condition
(\ref{eq:no-arb-const}) is met if $R$ takes values in $[0,1]$.

\subsection{CPS Basic Framework}

The modeling of the dependence between the default times of the
pool names is the key point in pricing financial products
depending in a non-linear way on the loss distribution. Typical
examples are CDO tranches and options on them.

We begin by briefly illustrating the
common Poisson shock framework (CPS), reviewed for example in Lindskog and McNeil (2003).

The occurrence of a default in a pool of names can be originated
by different events, either idiosyncratic or systematic. In the
CPS framework, the occurrence of the event number $e$, with
$e=1\dots m$, is modelled as a jump of a Poisson process
$N^{(e)}$. Notice that each event can be triggered many times.
Poisson processes driving different events are considered to be
independent.

The CPS setup assumes unrealistically that a defaulted name $k$
may default again. In the next section of the paper the try and
limit the number of defaults of each name to one. For now, we
assume that the $r$-th jump of  $N^{(e)}$ triggers a default event
for the name $k$ with probability $p_{r,k}^{(e)}$, leading to the
following dynamics for the single name default process $N_k$,
defined as the process that jumps each time name $k$ defaults:
\begin{equation}\nonumber
N_k(t) := \sum_{e=1}^m \sum_{r=1}^{N^{(e)}(t)} I^{(e)}_{r,k}
\end{equation}
where $I^{(e)}_{r,k}$ is a Bernoulli variable with probability
$\Qx{I^{(e)}_{r,k}=1}=p_{r,k}^{(e)}$. Under the Poisson assumption
for $N^{e}$ and the Bernoulli assumption for $I^{(e)}_{r,j}$ it
follows that $N_k$ is itself a Poisson process. Notice however
that the processes $N_k$ and $N_h$ followed by two different names
$k$ and $h$ are not independent since their dynamics is explained
by the same driving events.

The core of the CPS framework consists in mapping the single name
default dynamics, consisting of the dependent Poisson processes
$N_k$, into a multi-name dynamics explained in terms of {\it
independent} Poisson processes $\widetilde{N}_s$, where $s$ is a
subset (or ``cluster") of names of the pool, defined as follows.
\begin{equation}\nonumber
\widetilde{N}_s(t) = \sum_{e=1}^m \sum_{r=1}^{N^{(e)}(t)}
                 \sum_{s'\supseteq s} (-1)^{|s'|-|s|} \prod_{k'\in s'} I^{(e)}_{r,k'}
\end{equation}
where $|s|$ is the number of names in the cluster $s$. In a summation, $s \ni k$ means we are adding up across all clusters $s$ containing $k$, $k \in s$ means we are adding across all elements $k$ of cluster $s$, while $|s|=j$ means we are adding across all clusters of size $j$ and, finally, $s'\supseteq s$ means we are adding up across all clusters $s'$ containing cluster $s$ as a subset.

The non-trivial proof of the independence of $\widetilde{N}_s$ for
different subsets $s$ can be found in Lindskog and McNeil (2003).
Notice that a jump in a $\widetilde{N}_s$ processes means that all
the names in the subset $s$, {\em and only those names}, have
defaulted at the jump time. We denote by $\widetilde{\lambda}_s$
the intensity of the Poisson process $\widetilde{N}_s(t)$, and we
assume it to be deterministic for the time being, although we present extensions later.

\subsection{Cluster processes $(\widetilde{N}_s)$ as CPS building blocks}

One does not need to remember the above construction. All that
matters for the following developments are the independent
clusters default Poisson processes $\widetilde{N}_s(t)$. These
can be taken as fundamental variables from which (correlated)
single name defaults and default counting processes follow. The
single name dynamics can be derived based on these independent
$\widetilde{N}_s$ processes in the so-called fatal shock
representation of the CPS framework:
\begin{equation}\label{Neq}
N_k(t) = \sum_{s\ni\,k} \widetilde{N}_s(t), \ \ \ \mbox{or} \ \ \
d N_k(t) = \sum_{s\ni\,k} d \widetilde{N}_s(t),
\end{equation}
where the second equation is the same as the first one but in
instantaneous jump form. We now introduce the process
$\gpl_j(t)$, describing the occurrence of the simultaneous default
of any $j$ names whenever it jumps (with jump-size one):
\begin{equation}\label{Zeq}
\gpl_j(t):=\sum_{|s|=j} \widetilde{N}_s(t).
\end{equation}
Notice that each $\gpl_j(t)$, being the sum of independent Poisson
processes, is itself Poisson. Further, since the clusters
corresponding to the different $\gpl_1, \gpl_2, \ldots, \gpl_M$ do
not match, the $\gpl_j(t)$ are independent Poisson processes.

The multi-name dynamics, that is the default counting process
$\gpl_t$ for the whole pool, can be easily derived by carefully
adding up all the single name contributions.
\begin{equation}\nonumber
\gpl_t := \sum_{k=1}^M N_k(t) = \sum_{k=1}^M \sum_{s\ni\,k}
\widetilde{N}_s(t)
        = \sum_{k=1}^M \sum_{j=1}^M \sum_{s\ni\,k,|s|=j} \widetilde{N}_s(t)
        = \sum_{j=1}^M j \sum_{|s|=j} \widetilde{N}_s(t) ,
\end{equation}
leading to the relationship which links the set of dependent
single name default processes $N_k$ with the set of independent
and Poisson distributed counting processes $\gpl_j$:
\begin{equation}\label{bridge}
\sum_{k=1}^M N_k(t) = \sum_{j=1}^M j \gpl_j(t) =: \gpl_t
\end{equation}

Hence, the CPS framework offers us a way to consistently model the
single name processes along with the pool counting process taking
into account the correlation structure of the pool, which remains
specified within the definition of each cluster process
$\widetilde{N}_s$. Notice, however, that the $\gpl_t/M$ process is
not properly the re-scaled number of defaults $\defratenorm_t$,
since the former can increase without limit, while the latter is
bounded in the $[0,1]$ interval. We address this issue in Section
\ref{sec:avoiding:repeated} below, along with the issue of avoding repeated single names and cluster defaults.

\subsection{Equivalent formulation as compound Poisson process\label{sec:compound}}

One more way of looking at the $\gpl_t$ process is the compound
Poisson process, although the link with single name dynamics is
lost with this interpretation, since we cannot single out each
$\widetilde{N}_s$ given only the dynamics of $\gpl_t$. At any time
$t$ the process $\gpl_t$ has the same characteristic function as a
particular compound Poisson process. Consider the following
compound Poisson process
\[ X_t = \sum_{i=1}^{N_t} Y_i, \ \ \]
where $\lambda_j := \sum_{|s|=j} \widetilde{\lambda}_s$, $N$ is a
standard Poisson process with intensity $\lambda$ and the $Y_j$'s
are i.i.d random variables, independent of $N$, and with
distribution given by
\[ Y_j \sim \left\{ \begin{array}{cc}
 1 & \lambda_1/(\sum_{j=1}^M \lambda_j) \smallskip\\
 2 & \lambda_2/(\sum_{j=1}^M \lambda_j) \\ \vdots &  \\
 M & \lambda_M/(\sum_{j=1}^M \lambda_j)  \end{array} \right.
\]

If we define $\lambda := \sum_{j=1}^M \lambda_j$, then the compound
Poisson process $X_t$ has the same characteristic function, at all
times $t$, as the process $\gpl_t$.
The finite dimensional distributions of the two processes coincide
as well, so that substantially $\gpl_t$ and $X_t$ are the same process.
This is easily checked by writing the finite dimensional distributions
in terms of independent increments, while recalling that both $\gpl_t$
and $X_t$ have stationary independent increments.

Finally, we notice that also Di Graziano and Rogers (2005) in some
of their formulations obtain a compound Poisson process for the
loss distribution.

\subsection{Copula structure of default times}

The single name default dynamics in the CPS framework induces a
Marshall-Olkin copula type dependence between the first jumps of
the single name processes $N_j$. More precisely, if the random
default times $\{\tau_1,\dots,\tau_M\}$ of names $1,\ldots,M$ in
the pool are modeled as the first jump times of the single name
processes $N_1,\ldots,N_M$,
\begin{equation}\nonumber
\tau_k := \inf\{t\ge 0:N_k(t)>0 \},
\end{equation}
then Lindskog and McNeil (2003) show that the default times vector
is distributed according to a multi-variate distribution whose
survival copula is a $M$-dimensional Marshall-Olkin copula.


\section{Avoiding repeated defaults}\label{sec:avoiding:repeated}

In the above framework we have a fundamental problem, due to
repeated jumps of the same Poisson processes. Indeed, if the jumps are to be intepreted as defaults, this leads the above framework to unrealistic consequences. Indeed, repeated defaults
would occur both at the cluster level, in that a given cluster $s$ of
names may default more than once, as $\widetilde{N}_s$ keeps on
jumping, and at the single name level, since each name $k$ keeps
on defaulting as the related Poisson process $N_k$ keeps on
jumping. These repetitions would cause the default counting process
$\gpl_t$ to exceed the pool size $M$ and to grow without limit in
time.

There are two main strategies to solve this problem. Both take as
starting points the cluster repeated-default processes
$\widetilde{N}_s$ and then focus on different variables. They
can be summarized as follows.

{\bf Strategy 1 (Single-name adjusted approach)}. Force single
name defaults to jump only once and deduce clusters jumps
consistently.

{\bf Strategy 2 (Cluster adjusted approach)}. Force clusters to
jump only once and deduce single names defaults consistently.

The two choices have different implications, and we explore both
of them in the following, although we anticipate the second
solution is more promising.

If one gives up single names and clusters, and focuses only on the
default counting process and the loss (throwing away the
``bottom-up" interpretation), there is a third possible strategy
to make the default counting process above consistent with the
pool size:

{\bf Strategy 0 (Default-counting adjusted approach)}.  Modify the
aggregated pool default counting process so that this does not
exceed the number of names in the pool.

Strategy 0 addresses the problem of the CPS framework at the
default counting level. In the basic CPS framework, the link between the re-scaled pool
counting process $\gpl_t/M$, which can increase without limit, and
the re-scaled number of defaults $\defratenorm_t$, that must be
bounded in the $[0,1]$ interval, is not correct. This forbids in
principle to model $\defratenorm_t$ as $\gpl_t/M$. In the CPS
literature this problem is not considered usually. Lindskog and
McNeil (2003) for instance suppose that the default intensities of
the names are so small to lead to negligible ``second-default"
probabilities. If this assumption were realistic, this would allow
for adopting $\gpl_t/M$ as a model for $\defratenorm_t$ and
strategy 0 would not be needed. However, in our calibration
results in Brigo, Pallavicini and Torresetti (2006a) we find that
intensities are large enough to make repeated defaults
unacceptable in practice.

In Table~\ref{sec:table:notationcluster} we summarize the notation
we are going to adopt in the following.

\begin{table}\small
\begin{center}
\begin{tabular}{|c|c|c|c|c|}
\hline   & default process  & default proc  & default count proc &
total default
\\  & for cluster $s$ &  for
name $k$ &  for $j$ simult defaults & counting  \\ \hline   & & & & \\
 Repeated defaults & $\widetilde{N}_s$ & ${N}_k$ & ${\gpl}_j$ & ${\gpl}$ \\  & & & & \\
 Strategy 0 (GPL) & -- & -- & $ \gpl_j^0$    & $\min(\gpl, M)$  \\ &&&& \\
 Strategy 1 & $\widetilde{N}^1_s$ & $N^1_k$ & $\gpl^1_j$ & $\gpl^1$ \\ & & &  & \\
 Strategy 2 (GPCL) & $\widetilde{N}^2_s$ & $N^2_k$ & $\gpl^2_j$ & $\gpl^2$ \\& & &  & \\
\hline
\end{tabular}\caption{Notation for default processes according to
the different strategies}\label{sec:table:notationcluster}
\end{center}
\end{table}


\subsection{Default-counting adjustment: GPL
model (strategy 0)}\label{sec:firstgpl}

One possibility is to consider the pool counting process $\gpl_t$
merely as a driving process of some sort for the market relevant
quantities, namely the cumulated portfolio loss $\lossnorm_t$ and
the re-scaled number of defaults $\defratenorm_t$. This candidate
underlying process $\gpl_t$ is non-decreasing and takes
arbitrarily large values in time. The portfolio cumulated loss and
the re-scaled number of defaults processes are non-decreasing, but
limited to the interval $[0,1]$. Thus, we may consider a
deterministic non-decreasing function
$\psi:\mathbb{N}\cup\{0\}\rightarrow[0,1]$ and we define either
the counting or loss process as $\psi(\gpl_t)$. In Brigo,
Palavicini and Torresetti (2006a) we go for the former choice, by
capping the counting process coming from single name repeated
defaults, assuming
\begin{equation}\label{eq:gpl:cappingc}
\defratenorm_t := \psi_\defratenorm(\gpl_t) :=
\min(\gpl_t/M,1),
\end{equation} where $M>0$ is the number of
names in the portfolio, while in Brigo, Pallavicini and Torresetti
(2006b) we adopt the latter choice,
\begin{equation}\label{eq:gpl:cappingl}
\lossnorm_t := \psi_\lossnorm(\gpl_t) := \min(\gpl_t/M',1),
\end{equation}
where $1/M'$, with $M'\ge M>0$, is the minimum jump-size allowed for the
loss process, leading to more refined granularity solutions. The
quantity that is not modelled directly between $\defratenorm_t$
and $\lossnorm_t$ can be obtained from the one modelled directly
through explicit assumptions on the recovery rate. We discuss
recovery assumptions in general below, in
Section~\ref{sec:noabrconst-rec}.

This approach has the drawback of breaking the relationship
(\ref{bridge}) which links the single name processes $N_k$ with
the counting processes $\gpl_j$. We can still write the counting
processes as a function of the repeated default counting process
$\gpl_t$ under formula ($\ref{eq:gpl:cappingc}$):

\[ \gpl_j^0(t) = \int_0^t 1_{\{d \gpl_u=j , \gpl_{u^-} \le M-j\}} = \int_0^t  1_{\{\gpl_{u^-} \le M-j\}} d \gpl_j(u), \]

but we have clearly no link with single names.

This can be considered a viable approach, if we are interested only in the collective dynamics of the pool without considering its constituents, i.e. in
the aggregate loss picture typical of many ``top-down" approaches.

\subsection{Single-name adjusted approach (strategy 1)}\label{sec:ncheck-notcons}

In order to avoid repeated defaults in single name dynamics, we
can introduce constraints on the single name dynamics ensuring
that each single name makes only one default. Such constraints can be
implemented by modifying Equations (\ref{Neq}) in order to allow
for one default only. Given the same repeated cluster processes
$\widetilde{N}_s$ as before, we {\emph{define}} the new single
name default processes $N^1_k$ replacing $N_k$ as solutions of
the following modification of Equation~(\ref{Neq}) for the
original~$N_k$:
\begin{eqnarray}\label{Neq2}
d N^1_k(t) &:=& (1-N^1_k(t^-)) \sum_{s\ni\,k}
d\widetilde{N}_s(t)\\\nonumber
         &=& \sum_{s\ni\,k} d\widetilde{N}_s(t) \prod_{s\ni\,k} \ind{\widetilde{N}_s(t^-)=0}
\end{eqnarray}
{\bf Interpretation:} {\em This equation amounts to say that name
$k$ jumps at a given time if some cluster $s$ containing $k$ jumps
(i.e. $\widetilde{N}_s$ jumps) and if no cluster containing name
$k$ has ever jumped in the past.}

We can compute the new cluster defaults $\widetilde{N}^1_s$
consistent with the single names $N^1_k$   as
\begin{equation}\label{N1clust} d \widetilde{N}^1_s(t) = \prod_{j \in s} d N^1_j(t) \prod_{j \in
s^c} (1- d N^1_j(t))
\end{equation}
where $s^c$ is the set of all names that do not belong in $s$.

Now, we can use equation (\ref{bridge}) with the $N^1_k$
replacing the $N_k$, to calculate how the new counting processes
${\gpl}^1_j$ are to be defined in terms of the new single names
default dynamics:
\begin{eqnarray*}
\sum_{k=1}^M dN^1_k(t)
 &=& \sum_{k=1}^M (1-N^1_k(t^-)) \sum_{s\ni\,k} d\widetilde{N}_s(t)
 = \sum_{k=1}^M (1-N^1_k(t^-)) \sum_{j=1}^M \sum_{s\ni\,k,|s|=j} d\widetilde{N}_s(t)\\
 &=& \sum_{j=1}^M \sum_{|s|=j} d\widetilde{N}_s(t) \sum_{k\in s} (1-N^1_k(t^-))
 = \sum_{j=1}^M \sum_{|s|=j} d\widetilde{N}_s(t) \sum_{k\in s} \prod_{s'\ni k}
 \ind{\widetilde{N}_{s'}(t^-)=0}.
\end{eqnarray*}
This expression should match $ d {\gpl}^1(t):=\sum_j j \ d
{\gpl}^1_j(t)$, so that the counting processes are to be defined
as
\begin{equation}\label{Zeq2}
d {\gpl}^1_j(t) := \frac{1}{j} \sum_{|s|=j} d\widetilde{N}_s(t)
\sum_{k\in s} \prod_{s'\ni k} \ind{\widetilde{N}_{s'}(t^-)=0}
\end{equation}

The intensities of the above processes can be directly calculated
in terms of the density of the process compensator. We obtain by
direct calculation
\begin{eqnarray*}
h_{N^1_k}(t) &=& \prod_{s\ni\,k} \ind{\widetilde{N}_s(t^-)=0} \sum_{s\ni\,k} \widetilde{\lambda}_s(t) \\
h_{{\gpl}^1_j}(t) &=& \frac{1}{j} \sum_{|s|=j}
\widetilde{\lambda}_s(t)
               \sum_{k\in s} \prod_{s'\ni k} \ind{\widetilde{N}_{s'}(t^-)=0}
\end{eqnarray*}
where in general we denote by $h_X(t)$ the compensator density of
process $X$ at time $t$, referred to as ``intensity of $X$", and where $\widetilde{\lambda}_s$ is the intensity of the Poisson process $\widetilde{N}_{s'}$.

Given exogenously the repeated Poisson ``cluster" default building
blocks $\widetilde{N}_s$, the model $N^1_k, \widetilde{N}^1_s,
{\gpl}^1_j$ is a consistent way of simulating the single name
processes, the cluster processes and the pool counting process
from the point of view of avoiding repeated defaults. In
particular, we obtain $\defratenorm_t:= \sum_k {N}^1_k(t)/M =
{\gpl}^1_t/M \le 1$.

Notice, however, that the definition of $N^1_k$ in (\ref{Neq2}),
even if it avoids repeated defaults of single names, is not
consistent with the spirit of the original repeated cluster
dynamics.

Consider indeed the following example.

{\bf Begin Example.} {\em Consider two clusters $s = \{1,2,3\}$,
$z = \{3,4,5,6\}$. Assume no name defaulted up to time $t$ except
for cluster $z$, in that in a single past instant preceding $t$
names~$3,4,5,6$ (and only these names) defaulted together
($\widetilde{N}_{z}$ jumped at some past instant).
Now suppose at time $t$ cluster $s$ jumps, i.e. names~$1,2,3$
(and only these names) default, i.e.  $\widetilde{N}_{s}$ jumps
for the first time.

Question: Does name $2$ default at $t$?

According to our definition of $N^1_2$ the answer is yes, since no
cluster containing name $2$ has ever defaulted in the past. However,
we have to be careful in interpreting what is happening at {\em
cluster} level. Indeed, clusters $z$ and $s$ cannot both default
since this way name $3$ (that is in both clusters) would default
twice. So we see that the actual clusters default of this
approach, implicit in Equation~(\ref{N1clust}), do not have a
clear intuitive link with repeated cluster defaults
$\widetilde{N}_{s}$.}

 {\bf End Example.}

To simplify the parameters, we may assume the cluster intensities
$\widetilde{\lambda}_s$ to depend only on the cluster size
$|s|=j$. Then it is possible to directly calculate the intensity
of the pool counting process $\defrate = {\gpl}^1$ as
\begin{equation}\nonumber
h_{{\gpl}^1}(t) = \left(1-\frac{{\gpl}^1_{t^-}}{M}\right) \sum_j j {M
\choose j} \widetilde{\lambda}_j
\end{equation}
where $\widetilde{\lambda}_j$ is the common intensity of clusters
of size $j$.

We see that the pool counting process intensity $h_{{\gpl}^1}$ is
a linear function of the counting process $\defrate={\gpl}^1$
itself, as we can expect by general arguments for a pool of
{\emph{independent}} names (again with homogeneous intensities).
In such a pool default of one name does not affect the intensity
of default of other names, and the pool intensity is the common
homogeneous intensity times the number of outstanding names. Each
new default simply diminishes the pool intensity of one common
intensity value and the pool intensity is always proportional to
the number (fraction) of outstanding names $(1-\defratenorm)$.

\subsection{GPCL model: Cluster-adjusted
approach (strategy 2)}\label{sec:consistent-selfaffecting}

In the preceding sections we have seen that, if we are able to
model all the repeated cluster defaults $\widetilde{N}_s$, we are
able to describe the repeated default dynamics of both single
names and the pool as a whole. Indeed, by knowing all the
$\widetilde{N}_s$, we can directly compute the single name
processes $N_k$ and the aggregated counting processes $\gpl_j$ by
means of equations (\ref{Neq}) and (\ref{Zeq}).

In the previous section we have used the $\widetilde{N}_s$
exogenously as an engine to generate single name and aggregated
defaults. This avoids repeated defaults of single names and a
default rate exceeding 1, but is not consistent with the initial
intuitive meaning of the $\widetilde{N}_s$'s as repeated clusters
defaults.

The key to {\em consistently} avoid repeated cluster defaults (and
subsequently single names) is to track, when a cluster jumps,
which single-name defaults are triggered, and then force all the
clusters containing such names not to jump any longer.


We may formalize these points by introducing the process $J_s(t)$
defined as
\begin{equation}\nonumber
J_s(t) := \prod_{k\in s}\prod_{s'\ni
k}\ind{\widetilde{N}_{s'}(t)=0}
        = \prod_{s': \,s'\cap s\ne\emptyset}\ind{\widetilde{N}_{s'}(t)=0}
\end{equation}
The process $J_s(t)$ is equal to $1$ at starting time and it jumps
to $0$ whenever a cluster containing one element of $s$ jumps. Or
one may view the process $J_s$ as being one when none of the names
in $s$ have defaulted and $0$ when some names in $s$ have
defaulted. Notice that $J_s(t)=1$ implies
$\ind{\widetilde{N}_{s}(t)=0}$
 but not viceversa.

We now correct the cluster dynamics by avoiding repeated clusters
defaults. We define as new cluster dynamics the following:

\begin{equation}\label{Nbarequaltilde} d {\widetilde{N}^2}_s(t) = J_s(t^-) d
\widetilde{N}_s(t).
\end{equation}
{\bf Interpretation:} {\em every time a repeated cluster default
process $\widetilde{N}_s$ jumps, this is a jump in our
``no-repeated-jumps" framework only if no name contained in $s$
has defaulted in the past, i.e. if no cluster intersecting $s$ has
defaulted in the past.}

Once the clusters defaults are given, single name defaults follow
easily. We can change equation (\ref{Neq}) and define the single
name dynamics as
\begin{equation}\label{Neq3}
d N^2_k(t) := \sum_{s\ni\,k} d {\widetilde{N}^2}_s =
\sum_{s\ni\,k} J_s(t^-) d\widetilde{N}_s(t).
\end{equation}
Now, we can use equation (\ref{Zeq}) to see how the counting
processes $\gpl_j$ are to be re-defined in terms of our new cluster
dynamics  (\ref{Nbarequaltilde}). We obtain

\begin{eqnarray}\label{Zeq3}
d {{\gpl}^2}_j :=  \sum_{|s|=j} d {\widetilde{N}^2}_s =
\sum_{|s|=j} J_s(t^-) d\widetilde{N}_s(t).
\end{eqnarray}

The pool counting process reads

\begin{eqnarray}\label{Ztoteq3}
d {{\gpl}^2} =  \sum_{j=1}^M j \sum_{|s|=j} d {\widetilde{N}^2}_s = \sum_{j=1}^M j
\sum_{|s|=j} J_s(t^-) d\widetilde{N}_s(t).
\end{eqnarray}

If not for the cluster-related indicators $J_s(t^-)$,  ${{\gpl}^2}$ would be a generalized Poisson process. That is why we term the model
$N^2_k,  {\widetilde{N}^2}_s ,{{\gpl}^2}_j$ the Generalized Poisson Cluster-adjusted Loss model (GPCL).

Recall that we can always consider cluster dynamics as defined by
single name dynamics rather than directly. That is, we can define

\begin{equation}\label{Nbarclust} d {\widetilde{N}^2}_s(t) = \prod_{j \in s} d N^2_j(t) \prod_{j \in
s^c} (1- d N^2_j(t))
\end{equation}

This way the cluster $s$ defaults, i.e. ${\widetilde{N}^2}_s$
jumps (at most once), when (and only when) all single names in
cluster $s$ jump at the same time (first product), provided that
at that time no other name jumps (second product).

One can check that (\ref{Nbarclust}) and (\ref{Nbarequaltilde})
are indeed consistent if the single name dynamics is defined by
(\ref{Neq3}).

To appreciate how this second strategy formulation improves on the
first strategy, we consider again our earlier example.

{\bf Example (Reprise).} {\em Consider the same example as in
Section~\ref{sec:ncheck-notcons} up to the Question: ``Does name $2$
default at $t$?"

According to our definition of $N^2_2$ the answer is now NO, since
the cluster $z=\{3,4,5,6\}$, intersecting the $s$ currently
jumping (they both have name $3$ as element), has already defaulted in
the past. Thus we see a clear difference between strategies 1 and
2. With strategy 2 name $2$ does not default when $s$ jumps, with
strategy 1 it does. Notice that strategy 2 is more consistent
with the original spirit of the repeated cluster defaults
$\widetilde{N}_{s}$. Indeed, if cluster $z=\{3,4,5,6\}$ has
defaulted in the past (meaning that $\widetilde{N}_{z}$ has
jumped), $s = \{1,2,3\}$ should never be allowed to default, since
it is impossible that now ``exactly the names~$1,2,3$ default",
given that $3$ has already defaulted in $z$.}\\ {\bf End Example}

The intensities of the above processes can be directly calculated
as densities of the processes compensators. We obtain by direct
calculation, given that $J_s(t)$ is known given the information
(and in particular the  $\widetilde{N}_s$) at time $t$:
\begin{eqnarray}\label{intensity}
h_{N^2_k}(t) &=& \sum_{s\ni\,k} J_s(t^-) \widetilde{\lambda}_s(t)
\\   h_{\gpl^2_j}(t) &=& \sum_{|s|=j} J_s(t^-)
\widetilde{\lambda}_s(t)
\end{eqnarray}

\begin{remark} {\bf (Self-affecting features ).}
Notice that in the GPCL model the single name intensities $h_{N^2_k}(t)$ are stochastic, since they depend on the process $J_s$. Moreover, the single name intensities are affected by the loss process. In particular, the intensity of a single-name jumps when one of the other names jumps. Consider for example a
name $k$ that has not defaulted by $t$, with intensity
$h_{N^2_k}(t)$, and one path where there are no new defaults
until $t'> t$, when name $k'$ defaults. Now all clusters $s$
containing $k'$ have $J_{s}(t')=0$ so that
\[ h_{N^2_k}(t')
= \sum_{s\ni\,k} J_s(t'^-) \widetilde{\lambda}_s(t') =
\sum_{s\ni\,k} J_s(t^-) \widetilde{\lambda}_s(t') - \sum_{s
\supseteq\,\{k,k'\}} J_s(t^-) \widetilde{\lambda}_s(t')\]

We see that the the $k$-th name intensity reduces when $k'$ defaults,
and it reduces of the second summation in the last term.

At first sight this is a behaviour that is not ideally suited to intensities. For example, looking at the loss feedback present in the default intensities of Hawkes-processes (see Errais, Giesecke and Goldberg (2006) for Hawkes processes applied to default modeling), one sees that intensities are self-exciting, in that they {\em increase} when a default arrives. As soon as one name defaults, the intensities of the pool jump up, as is intuitive. However, Errais, Giesecke and Goldberg (2006) (but also Sch\"{o}nbucher (2005) and others) assume there is only one default at a time. We are instead assuming there may be more than one default in a single instant. Therefore the self-exciting feature is somehow built in the fact that more than one name may default at the same instant. In other terms, instead of having the intensity of default of a related name jumping up of a large amount, implying that the name could default easily in the next instants, we have the two names defaulting together. From this point of view cluster defaults embed the self-exciting feature, although in an extreme way.
\end{remark}

The best way to summarize our construction is through the three
equations defining respectively cluster defaults, single name
defaults and default counting processes:

\[ d {\widetilde{N}^2}_s(t) = J_s(t^-) d \widetilde{N}_s(t),  \  \ \boxed{ d N^2_k(t) := \sum_{s\ni\,k} d {\widetilde{N}^2}_s(t),
\ d\gpl^2_j(t) := \sum_{|s|=j} d {\widetilde{N}^2}_s(t) }\]

Notice that once the new cluster default processes
${\widetilde{N}^2}_s$ are properly defined, single name and
default counting processes follow immediately in what is indeed
the only possible relationships that make sense for connecting
clusters fatal shocks to single name defaults and to default
counting processes. With our particular choice for the cluster
defaults ${\widetilde{N}^2}_s$ dynamics we start from the repeated
cluster defaults $\widetilde{N}_s$ dynamics and correct it to
avoid repeated defaults at a cluster level. Then everything
follows for default counting and single names.

We may also write the cluster intensities as

\[ h_{{\widetilde{N}^2}_s}(t)
=  J_s(t^-) \widetilde{\lambda}_s(t) =: \bar{\lambda}_s(t) \]

Notice that this strongly reminds us of what we do with Poisson
(or more generally Cox) processes to model single name defaults.
The default time $\tau_k$ of the single name $k$ is modeled as the
first jump of a Poisson process with intensity $\lambda_k(t)$, and
then the process is killed after the first jump in order to avoid
repeated defaults. This way the intensity $\bar{\lambda}_k(t)$ of
the default time $\tau_k$ is
\[ \bar{\lambda}_k(t) =  1_{\{\tau_k > t  \}} \lambda_k(t) \]
What we do is similar for clusters: we start from clusters with
repeated jumps $\widetilde{N}_s$ and then we kill the repeated
jumps through an indicator $J_s(t)$, replacing the simpler
indicator $1_{\{\tau_k > t  \}}$ of the single-name case.

If, as before, {\em we assume the cluster intensities
$\widetilde{\lambda}_s$ to depend only on the cluster size,
$\widetilde{\lambda}_s= \widetilde{\lambda}_{|s|}$}, it is
possible to directly calculate the intensity of the pool counting
process $\gpl^2(t) := \sum_j j \gpl^2_j(t)$. We obtain
\begin{equation}\nonumber
h_{\gpl^2}(t) = \sum_j j \, {M-\gpl^2_{t^-} \choose j}
\widetilde{\lambda}_j
\end{equation}
where $\widetilde{\lambda}_j$ is the common intensity of clusters
of size $j$. The pool counting process intensity is a non-linear
function of the counting process, taking into account the
co-dependence of single name defaults.


\subsection{Comparing models in a simplified
scenario}\label{htoverh0}

It is interesting to compare the relationships between the pool
counting process $\defrate_t$ and its intensity across the
different formulations we considered above.

Here, we summarize the approaches shown above {\em in the case
cluster intensities depend only on the cluster size, $\lambda_s =
\lambda_{|s|}$}.
\begin{figure}
\begin{center}
\includegraphics[width=0.7\textwidth]{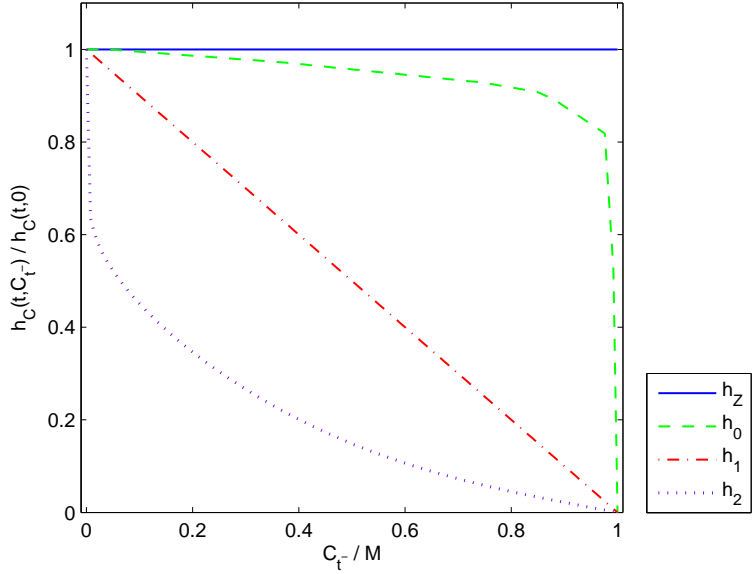}
\end{center}
\caption{\label{fig:intensity} The relationships between the pool
counting process $C_{t^-}/M$ and its intensity ratio
$h_\defrate(t;C_{t^-})/h_\defrate(t;C_{t^-}=0)$ in the four different
cases summarized in Section~\ref{htoverh0}. The cluster intensities
for the GPL and GPCL models are listed in the rightmost columns of
the two panels of Table \ref{tab:param}.}
\end{figure}

\begin{enumerate}
\item {\bf Repeated defaults}.
  The counting process can increase without limit, as
  implictly done in Lindskog and McNeil (2003).
  \begin{equation}\nonumber
    \defrate_t = \gpl_t, \ \
    h_\gpl(t) = h_\defrate(t) = \sum_{j=1}^M j\, {M \choose j} \widetilde{\lambda}_j(t)
  \end{equation}
\item {\bf Strategy 0}.
  The counting process is bounded by the mapping
  $\psi(\cdot) := \min(\cdot,M)$, as in Brigo, Pallavicini and
  Torresetti (2006a). This is the {\bf Generalized Poisson Loss (GPL) model}.
  \begin{equation}\nonumber
    \defrate_t = \min(\gpl_t, M),  \ \
    h_0(t) := h_\defrate(t) =
     \sum_{j=1}^M \min(j,(M-\gpl_{t^-})^+) {M \choose j} \widetilde{\lambda}_j(t)
  \end{equation}
\item {\bf Strategy 1}.
  The counting process is bounded by forcing each single
  name to jump at most once. A dynamics, leading to a similar
  form of the intensity, is considered also in Elouerkhaoui (2006).
  \begin{equation}\nonumber
    \defrate_t = {\gpl}^1_t ,  \ \
    h_1(t) := h_{\defrate}(t) =
     \left(1-\frac{{\gpl}^1_{t^-}}{M}\right) \sum_{j=1}^M j\, {M \choose j} \widetilde{\lambda}_j(t)
  \end{equation}
\item {\bf Strategy 2}.
  The counting process is bounded by forcing clusters dynamics
  to give raise to  at most one jump in each single name.
  This is the {\bf Generalized Poisson Cluster Loss (GPCL) model}.
  \begin{equation}\nonumber
    \defrate_t = \gpl^2_t, \ \
    h_2(t) := h_{\defrate}(t) =
     \sum_{j=1}^M j\, {M-\gpl^2_{t^-} \choose j} \widetilde{\lambda}_j(t)
  \end{equation}
\end{enumerate}

In Figure \ref{fig:intensity} we plot
$h_\defrate(t;C_{t^-})/h_\defrate(t;C_{t^-}=0)$ against $C_{t^-}/M$ in the
four cases. The cluster intensities $\widetilde{\lambda}_j$ for the
first and the third model are not relevant, since their influence
cancels taking the ratio. The cluster intensities for the second
and the fourth model are calibrated against the 10-year
DJi-TRAXX tranche and index spreads on October 2, 2006 (see Table
\ref{tab:param}).

Notice, further, that for any choice of the cluster intensities
the pool intensities are monotonic non-increasing functions of the
pool counting process, not explicitly depending on time.

\section{Beyond GPL: The GPCL model calibration}

In Brigo, Pallavicini and Torresetti (2006a) the GPL basic model
$\defrate_t = \min(\gpl_t, M)$ is calibrated to the index and its
tranches for several maturities. Here we try instead the richer
GPCL model $\defrate_t = \gpl^2_t$ introduced above, allowing us
in principle to model also cluster and single name defaults
consistently. However, the GPCL model can hardly be managed
without simplifying assumptions. In the following we assume again
that the cluster intensities $\widetilde{\lambda}_s$ depend only
on the cluster size~$|s|$. Moreover, as with the basic GPL model,
we try calibration of multi-name products only, such as credit
indices and CDO tranches, leaving aside single name data for the
time being. Indeed, with respect to our earlier paper in Brigo
Pallavicini and Torresetti (2006a), we focus only on the
improvement in calibration due to using a default counting process
whose intensity has a clear interpretation in terms of default
clusters. This will allow us, in further work, to include single
names in the picture, since our GPCL framework allows us to do so
explicitly.

The recovery rate is considered as a deterministic constant and
set equal to $R = 40\%$. Thus, the underlying driving model definition
is
\begin{equation}\nonumber
\defrate_t := \gpl^2(t) = \sum_{j=1}^M j\, \gpl^2_j(t)
,\ \ \mbox{where} \ \ \ d \gpl^2_j(t) \sim {\rm
Poisson}\left({M-\gpl^2_{t^-} \choose j} \widetilde{\lambda}_j(t) dt
\right)
\end{equation}
while the pool counting and loss processes are defined as
\begin{eqnarray*}
d \defratenorm_t &:=&  d \gpl^2_t/M\\
d \lossnorm_t &:=& (1-R)\, d \gpl^2_t/M
\end{eqnarray*}

In the following sections we first discuss the numerical issues
concerning calibration, and, then, we show some model calibration
results.

\subsection{Numerical issues concerning calibration}

Given our recovery assumption, the prices of the products to be calibrated, presented in the appendix, depend only on knowledge of the probability distribution of the pool counting process $C_t$. Thus, our main issue is to
calculate this law as fast as possible. When dealing with dynamics
derived from Poisson processes, there are different available
calculation methods, depending on the structure of the
intensities.

If the intensity does not depend on the process itself, or it does
only in a simple way, then the probability distribution can be
derived by means of Fast Fourier inversion of the characteristic
function, when the latter is available in closed form. This method
is described and used for the GPL model in Brigo, Pallavicini and
Torresetti (2006a,b). Again, the GPL process is based on
the driver $\gpl$, that can be interpreted also as a compound Poisson
process, as we have seen in Section \ref{sec:compound}. In general
the probability distributions of compound Poisson processes can be
calculated in closed form if the i.i.d. jump amplitudes have a
discrete-valued distribution. Consider the compound Poisson
process defined in Section \ref{sec:compound}. It is possible to
find a relationship, known as Panjer recursion, between the
probability densities $p_{X_t}(n)$ and $p_{X_t}(n-1)$ as done in
Hess et al. (2002).

However, with the GPCL model, the dependence of the intensity of
the pool counting process on the process itself prevents us either
to calculate the relevant characteristic function in closed form
or to use the Panjer method.

Our choice then is to explicitly calculate the forward Kolmogorov
equation satisfied by the probability distribution
$p_{\gpl^2_t}(x) = \Qx{\gpl^2_t = x}$, namely
\begin{equation}\nonumber
\frac{d}{dt}p_{\gpl^2_t}(x) = \sum_{y=0}^M A_t(x,y)
p_{\gpl^2_t}(y)
\end{equation}
where the transition rate matrix $A_t =
(A_t(x,y))_{x,y=0,\ldots,M}$ is given by \begin{equation}\nonumber
A_t(x,y) := \lim_{\Delta t\rightarrow 0}
\frac{\Qx{\gpl^2_{t+\Delta t}=x|\gpl^2_t=y} }{\Delta t}
          =  {M- y \choose x-y} \widetilde{\lambda}_{x-y}(t)
\end{equation}
for $x > y$,
\begin{equation}\nonumber
A_t(y,y) := \lim_{\Delta t\rightarrow 0}
\frac{\Qx{\gpl^2_{t+\Delta t}=y|\gpl^2_t=y}-1 }{\Delta t}
= -\sum_{j=1}^{M-y} {M-y \choose j} \widetilde{\lambda}_j(t).
\end{equation}
for $x=y$, and zero for $x< y$.

In matrix form we write
\[ \frac{d}{dt} \widehat{\pi}_t  = A_t \widehat{\pi}_{t} , \  \  \   \ \ \widehat{\pi}_{t} := \left[ \begin{array}{ccccc} p_{\gpl^2_t}(0)
& p_{\gpl^2_t}(1)& p_{\gpl^2_t}(2)& \ldots & p_{\gpl^2_t}(M)
\end{array} \right]'
\]
whose solution is obtained through the exponential matrix,
\[ \widehat{\pi}_t = \exp\left(\int_0^t A_u du\right) \widehat{\pi}_0, \ \ \widehat{\pi}_0 = [1 \ 0 \ 0 ... \ 0 ]'.  \]
Matrix exponentiation can be quickly computed with the Pad\'e
approximation (see Golub and Van Loan (1983)), leading to a closed
form solution for the probability distribution $p_{\defrate_t} =
\widehat{\pi}_t$ of the pool counting process $\defrate_t$. This
distribution can then be used in the calibration procedure.

\subsection{GPCL model detailed Calibration procedure}

If we define the cumulated cluster intensities as
\[
\widetilde{\Lambda}_j(t) =  \int_0^t \widetilde{\lambda}_j(u) \,du .
\]

then the entries of the matrix undergoing exponentiation in determining the default counting distribution are given by
\begin{eqnarray*}
{\rm for~} x > y {\rm :~}  \int_0^t A_u(x,y) du = {M-y \choose x-y}  \widetilde{\Lambda}_{x-y}(t)  \\
{\rm for~} x=y {\rm :~}     \int_0^t A_u(y,y) du = -
\sum_{j=1}^{M-y} {M-y \choose j} \widetilde{\Lambda}_{j}(t) .
\end{eqnarray*}

We assume the $\widetilde{\Lambda}_{j}$ to be piecewise linear in time, changing their values at payoff maturity dates. We use  $\widetilde{\Lambda}_{j}$
as calibration parameters.   We have $bM$ free calibration parameters, if we consider $b$ maturities. Notice that many $\widetilde{\Lambda}_j(t)$ will be equal to zero for all
maturities, meaning that we can ignore their corresponding
counting  process $\gpl^2_j(t)$. One can think of deleting all
the modes with jump sizes having zero intensity and keep only the nonzero intensity ones. Call
 $\alpha_1 < \alpha_2 < ... < \alpha_n$ the jump sizes with nonzero intensity. Then one renumbers progressively the intensities according to the nonzero
increasing $\alpha$: $\gpl^2_j$ becomes the jump of a cluster of
size $\alpha_j$.

The calibration procedure for GPCL is implemented using the $\alpha_j$ in the same way as in Brigo, Pallavicini and Torresetti (2006a) for the GPL
model. As concerns the GPCL intensities, in the tables we display ${M \choose \alpha_j} \widetilde{\Lambda}_{j}$, i.e. we multiply a cluster cumulated intensity for a given cluster size for the number of clusters with that size at time $0$.

We also calibrate the GPL model, for comparison. In this paper we denote the GPL cumulated intensities for the $\alpha_j$ mode by ${\Lambda}_{j}^0$, which reads, using the link with repeated defaults, as  ${\Lambda}_{j}^0 = {M \choose \alpha_j} \widetilde{\Lambda}_j$. Given the arbitrary a-posteriori capping procedure in GPL, these $\widetilde{\Lambda}_j$ are not to be interpreted as cluster parameters, the only actual parameters being the  ${\Lambda}_{j}^0$ directly, and they are to be interpreted as merely describing the pool counting process dynamic features.

More in detail, the optimal values for the
amplitudes $\alpha_j$ in GPCL are selected, by adding non-zero
amplitudes one by one, as follows, where typically $M=125$:
\begin{enumerate}
\item set $\alpha_1=1$ and calibrate $\widetilde{\Lambda}_1$;
\item add the amplitude $\alpha_2$ and find its best integer value
      by calibrating the cumulated intensities $\widetilde{\Lambda}_1$
      and $\widetilde{\Lambda}_2$, starting from the previous value for
      $\widetilde{\Lambda}_1$ as a guess, for each value of $\alpha_2$
      in the range $[1,125]$,
\item repeat the previous step for $\alpha_i$ with $i=3$ and so on,
      by calibrating the cumulated intensities
      $\widetilde{\Lambda}_1,\dots,\widetilde{\Lambda}_i$, starting from the previously
      found $\widetilde{\Lambda}_1,\dots,\widetilde{\Lambda}_{i-1}$ as initial
      guess, until the calibration error is under a pre-fixed threshold
      or until the intensity $\widetilde{\Lambda}_i$ can be considered negligible.
\end{enumerate}
The objective function $f$ to be minimized in the calibration is
the squared sum of the errors shown by the model to recover the
tranche and index market quotes weighted by market bid-ask
spreads:
\begin{equation}\label{eq:calibrerror}
f(\alpha,\widetilde{\Lambda}) =
 \sum_i \epsilon_i^2, \ \ \epsilon_i =
  \frac{x_i(\alpha,\widetilde{\Lambda})-x_i^{\rm Mid}}{x_i^{\rm Bid}-x_i^{\rm Ask}}
\end{equation}
where the $x_i$, with $i$ running over the market quote set, are
the index values $\cds_0$ for DJi-TRAXX index quotes, and either
the index periodic premiums $\cdstr_0$ or the upfront premium
rates $\upfronttr$ for the DJi-TRAXX tranche quotes, see the
appendix for more details.

\subsection{Calibration results}

The calibration data set is the DJi-TRAXX main series on the run
on October, 2 2006. In Tables \ref{input:discount} and
\ref{input:tranche} we list the discount interest rates, the CDO
tranche spreads and the credit index spreads.

We calibrate three methods against such data set and we compare
the results. They are listed in the tables.
\begin{enumerate}
\item The implied expected tranched loss method (hereafter ITL)
described in  Walker
(2006) or in Torresetti, Brigo and Pallavicini (2006).
      It is a method which allows to check if arbitrage
      opportunities are present on the market by implying expected
      tranched losses satisfying basic no-arbitrage requirements.
\item The GPL model described in Brigo, Pallavicini and Torresetti
(2006a) and summarized above, i.e. $\defrate_t = \min(\gpl_t,M)$ with $\gpl$ as in
(\ref{bridge}) (referred to before as strategy 0).
      Such model, due to the capping feature, is not compatible with any of the previously described single-name dynamics avoiding repeated defaults.
\item The GPCL model described in the present paper (strategy 2),
which represents an articulated solution to the repeated
defaults problem. We implement the simplified version with cluster
intensity $\widetilde{\lambda}_s$ depending only on cluster size
$|s|$.
\end{enumerate}

First, we check that there are no arbitrage opportunities on
October, 2 2006, by calibrating the ITL method. The calibration is
almost exact and in Table~\ref{tab:itl} we show the expected
tranched losses implied by the method, which we can use as
reference values when comparing the other two models.

Then we calibrate the GPL and GPCL models, and we obtain the
calibration parameters presented in Table~\ref{tab:param}, while
the expected tranched losses implied by these two models are
included in Table~\ref{tab:itl}. We point out that this is a joint
calibration across tranche seniority and maturity, since we are
calibrating all and every tranche and index quote with a single
model specification.  When looking at the outputs of the
calibrated models on the different maturities, we see that both
our models perform very well on maturities of 3 years, 5 years and
7 years, for which the calibration error is within the bid-ask spread.
The 10 year maturity quotes are more difficult to recover, but both
models are close to the market values, as we see from the left panel
of Table~\ref{tab:cal}. Notice, however, that the GPCL model has a
lower calibration error ($10\%-20\%$ better).

\begin{figure}
\begin{center}
\begin{tabular}{cc}
\includegraphics[width=0.48\textwidth]{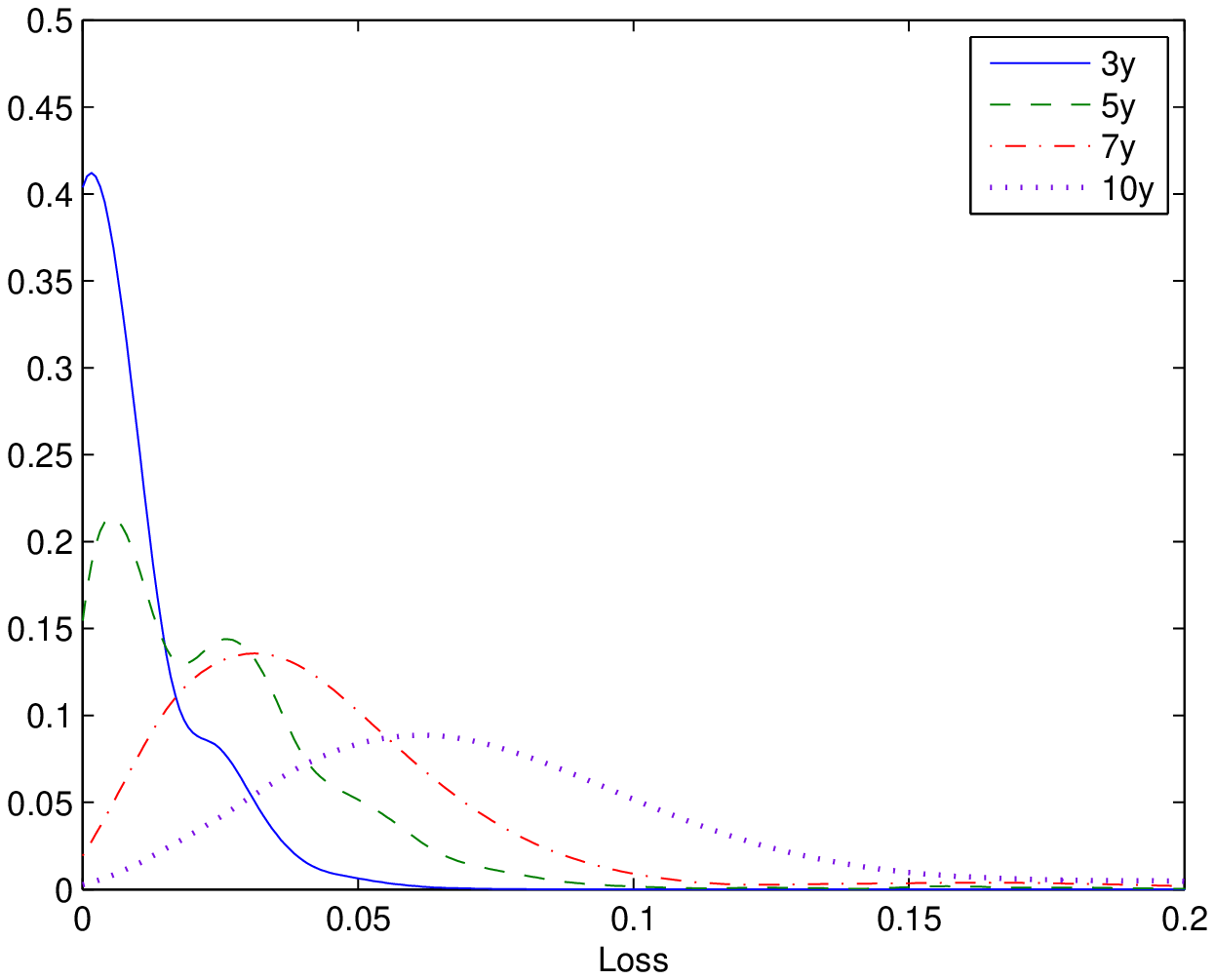} &
\includegraphics[width=0.48\textwidth]{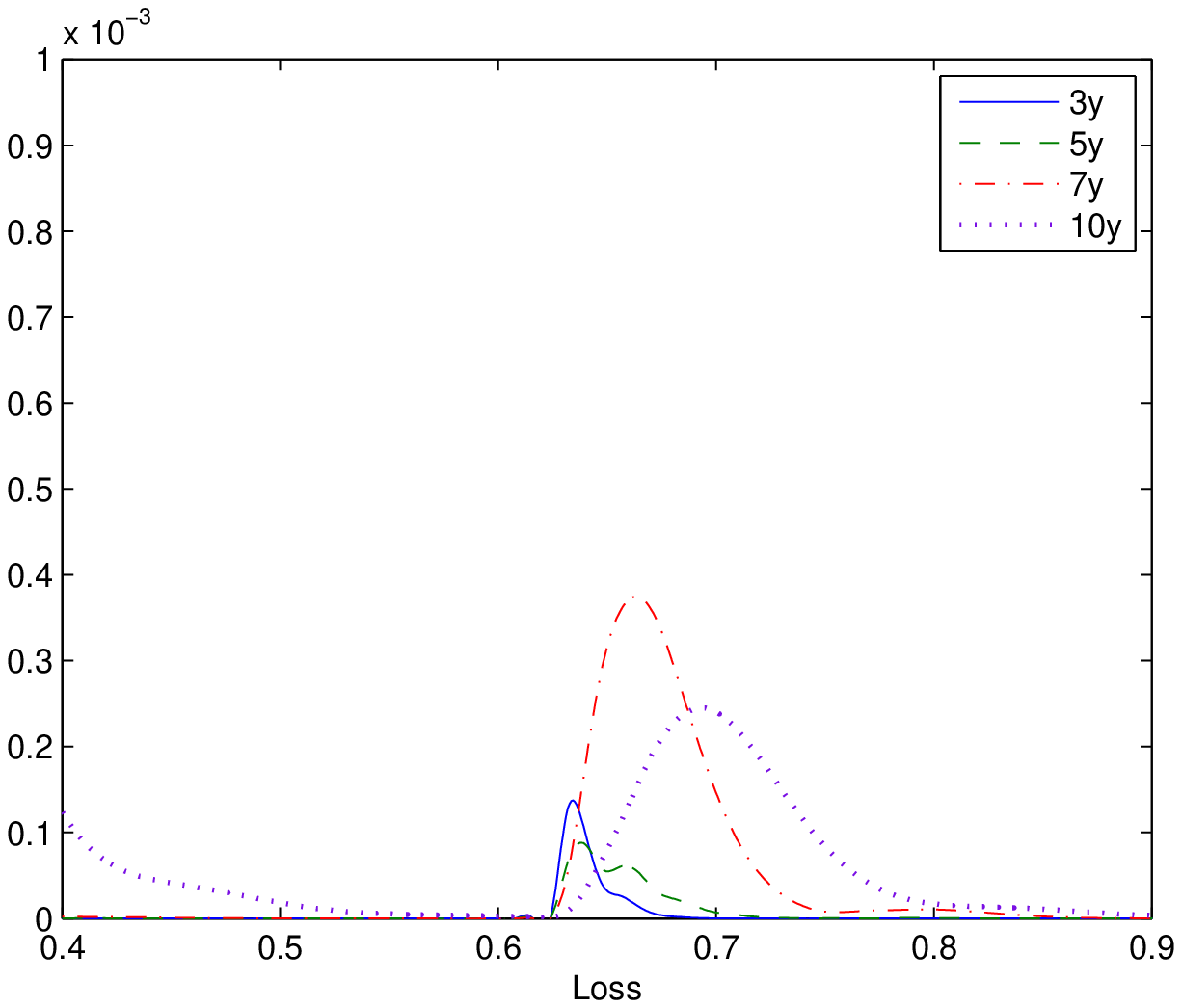} \\
\includegraphics[width=0.48\textwidth]{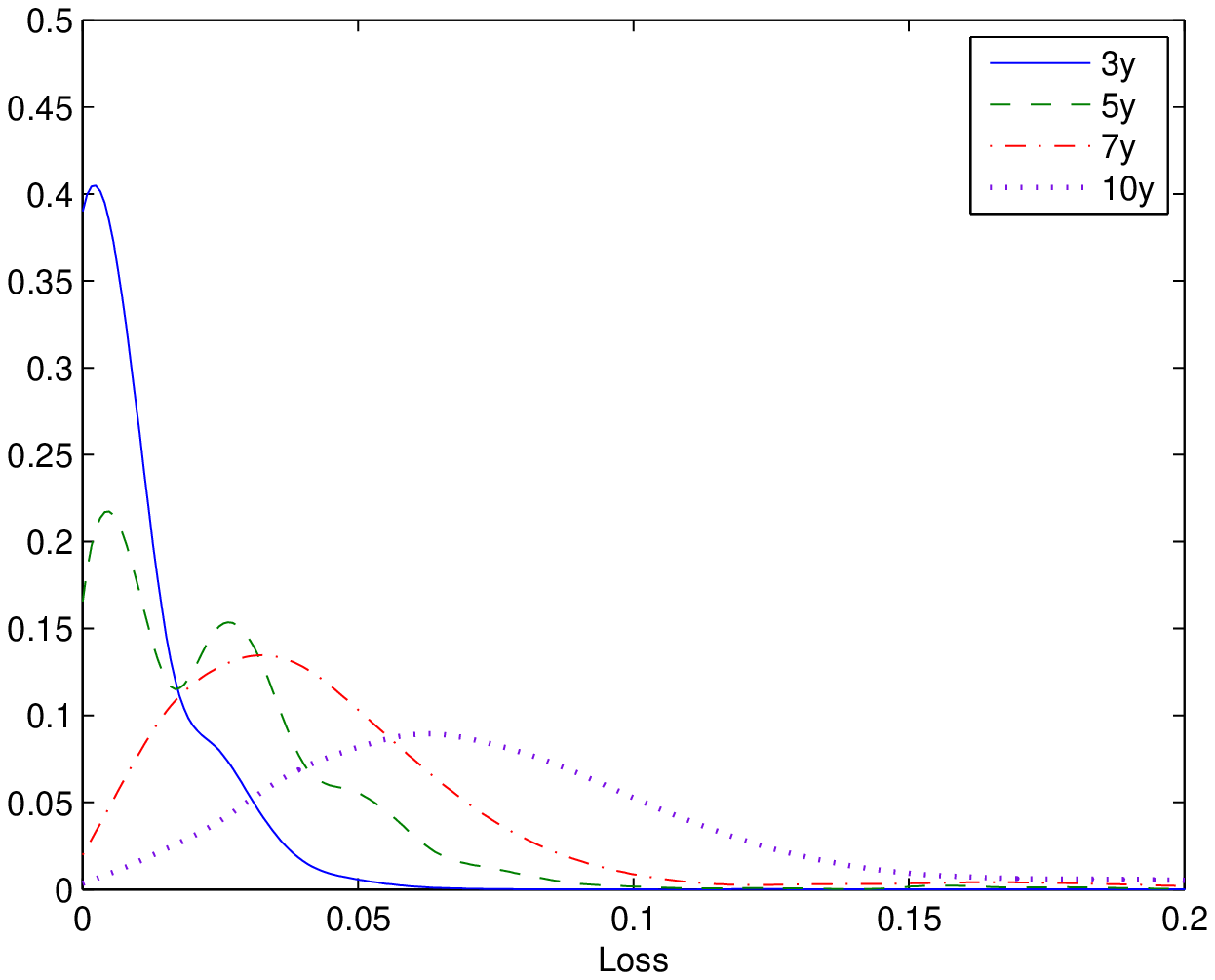} &
\includegraphics[width=0.48\textwidth]{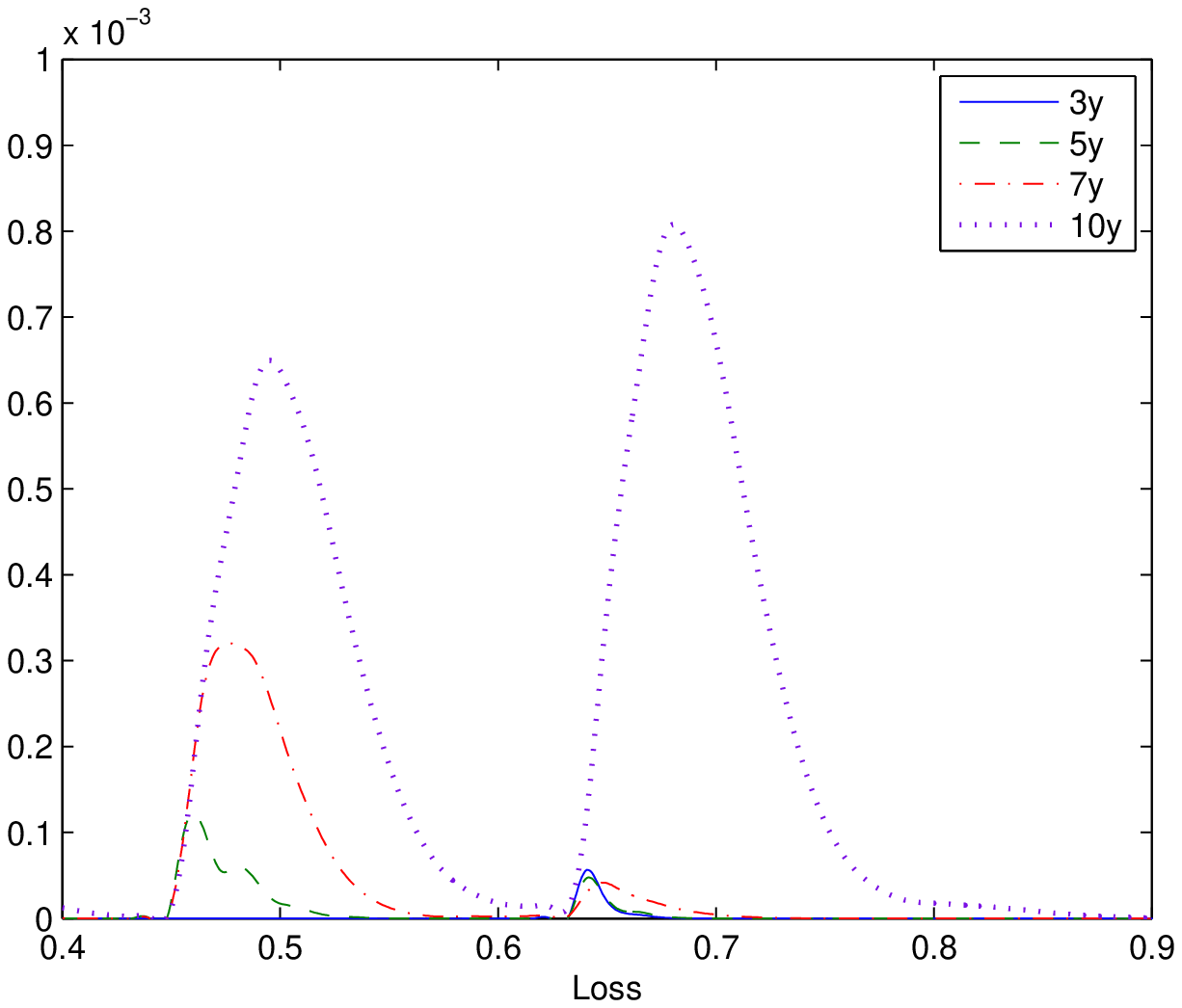} \\
\end{tabular}
\end{center}
\caption{\label{fig:prob} Loss distribution evolution of the GPL
model (upper panel) and of the GPCL model (lower panel) at all the
quoted maturities up to ten years, drawn as a continuous line.}
\end{figure}

The probability distributions implied by the two dynamical models
are similar at gross-grain view, as one can see in Figure
\ref{fig:prob}, but they differ if we
observe the fine structure. Indeed, the tails of the two
distributions show different bumps. The GPCL model shows a more
complex pattern, and, as one can see from Table \ref{tab:param},
its highest mode is the maximum portfolio loss, while the GPL
model has a less clear tail configuration.

We also apply the ITL, GPL and GPCL methods to the CDX index and
tranches (see Table~\ref{input:tranchecdx} for market quotes),
following the same procedure used for the DJi-TRAXX above.
We find better results, that are summarized in Table~\ref{tab:paramcdx}
and in the right panel of Table~\ref{tab:cal}.

\section{Model Extensions}\label{sec:extensionsgpcl}

In this final section we hint at possible extensions of the basic
model to account for more sophisticated features.

\subsection{Spread dynamics}

The valuation of credit index forward contracts or options
maturing at time $T=T_a$ requires the calculation of the index
spread at those future times, which in turn depends on the default
intensity evolution. Consider, for instance, the case of
deterministic interest rates (or more generally interest rates
independent of defaults) for an index whose default leg protects
against losses in the index pool up to time $T_b$ and where the
spread premium payments occur at times $T_{1},T_{2},\ldots, T_b$.
We have the spread expression at $T_a$ as
\[
\cds_T = \frac{\int_T^{T_b} D(T,t) \,\Ex{T}{h_\lossnorm(t)} dt}
           {\sum_{i=1}^b \, \delta_i D(T,T_i)
           \left(1-\defratenorm_T-\int_T^{T_i} \Ex{T}{h_\defratenorm(t)} dt \right) \ind{T_i>T}}
\]
where $h_\lossnorm(t)$ is the default intensity of the cumulated
portfolio loss process and $h_\defratenorm(t)$ is the default
intensity of the re-scaled default counting process $\defratenorm$
(see for example Brigo, Pallavicini e Torresetti (2006a), or the Appendix, for a
detailed description of credit index contracts) and $D(s,t)$ is
the discount factor, often assumed to be deterministic, between
times $s$ and $t$.

The GPCL model presented in the previous sections has single-name
and default counting intensities given by equations
(\ref{intensity}). These intensities depend on which names have
already defaulted. The dynamics of the index $S_t$ (spread
dynamics) can be enriched by more sophisticated modelling of the
default intensities $h_\lossnorm(t)$ and $h_\defratenorm(t)$, by
explicitly adding stochasticity to the Poisson intensities
$\widetilde{\lambda}_j(t)$, e.g. resorting to the Gamma, scenario
or CIR extensions of the model seen above.

\subsection{Spread dynamics through Gamma intensity}

Assume now that the cumulated clusters intensities
$\widetilde{\Lambda}_s(t):=\int_0^t\,\widetilde{\lambda}_s(u)du$
are distributed at any time $t$ according a Gamma distribution,
i.e.
\[ \widetilde{\Lambda}_s(t) \sim \Gamma(k_s(t), \theta_s) \]
where $k>0$ is the shape parameter and $\theta > 0$ is the scale
parameter in the Gamma distribution. These gamma processes are
assumed to be independent of the exponential random variables
triggering the jumps in the $\widetilde{N}_s$ Poisson
processes. The Gamma choice is convenient because it does not alter
the tractability of the basic model. See Brigo, Pallavicini e
Torresetti (2006b) for a Gamma GPL implementation.

The Gamma distribution assumption for $\widetilde{\Lambda}_s(t)
\sim \Gamma(k_s(t),\theta_s)$ at every time $t$ is consistent with
a Gamma process assumption for $\widetilde{\Lambda}_s(t)$, whose
distribution is controlled by both parameters $k$ and $\theta$.
The time constant $\theta$ allows for little flexibility in the
variance term-structure of the process. In Brigo, Pallavicini and
Torresetti (2006b) we improve the model in this respect, by
introducing a piecewise Gamma GPL process extension.

\subsection{Spread dynamics through  CIR intensity}

A different and possibly more interesting extension is to model
the cluster intensities according to a Cox Ingersoll Ross (CIR)
process
\begin{equation}\nonumber
d\widetilde{\lambda}_s(t) = k_s (\theta_s -
\widetilde{\lambda}_s(t)) dt
                      + \sigma_s \sqrt{\widetilde{\lambda}_s(t)} d W_s
\end{equation}
with $2 k_s \theta_s > \sigma_s^2$. These CIR processes are
assumed to be independent of the exponential random variables
triggering the jumps in the $\widetilde{N}_s$ Poisson processes.

With respect to the case of deterministic cluster intensities, the
model tractability is preserved, due to the closed form results
which can be derived. Alternatively, jump diffusion JCIR
intensities can be considered, maintaining tractability.

\subsection{Spread dynamics through  Scenario intensity}

A different extension is as follows. By taking scenarios on the
clusters intensities we may easily extend our basic model. In this
model we assume the intensities in all the clusters to take
different scenarios with different probabilities. Indeed, assume
now that the (possibly time varying) intensities
$\widetilde{\lambda}_s$ are indexed by a random variable $I$
taking values $1,2,\ldots,m$ with (risk-neutral) probabilities
$q_1, q_2, \ldots, q_m$: $\widetilde{\lambda}_s^I$ is then a
random intensity for the $s$-th cluster process, depending on $I$.
The related Poisson process is denoted by $\widetilde{N}_s^I$. $I$
is assumed to be independent of the exponential random variables
triggering the jumps of the Poisson processes. Conditional on
$I=i$, the intensity of the process $\widetilde{N}^I_s$ is
$\widetilde{\lambda}_s^i$. This formulation does not spoil
analytical tractability, since all the expected values can be
calculated as a linear combination of conditional expectations.

\subsection{Recovery dynamics}\label{sec:noabrconst-rec}

We introduced in~(\ref{eq:afc}), reported below here, the notion
of recovery at default $\rec_t$:
\begin{equation}
\label{eq:afc:bis}    d \lossnorm_t =  (1-\rec_t) d \defratenorm_t \ \   \ \ ( \mbox{or, more precisely} \ \ \  \lossnorm_t = \int_0^t (1-\rec_u) d \defratenorm_u) .
\end{equation}

Now we specify more about this notion. In general, for ease of
computation, we assume $\rec_t$ to be a
${\cal G}_t$-adapted and left-continuous (and hence predictable) process taking values in the interval $[0,1]$. On predictability of the recovery process see also Bielecki and Rutkowski (2001).
Here ${\cal G}_t$ denotes the filtration consisting of
default-free market information and of the default-count
monitoring up to time $t$. This implies in particular, given
(\ref{eq:afc:bis}), that the loss $\lossnorm_t$ is ${\cal
G}_t$-adapted too, as is reasonable. We noticed earlier that the no-arbitrage
condition (\ref{eq:no-arb-const}) is met if $R$ takes values in
$[0,1]$. Equation (\ref{eq:afc:bis}) leaves us with the freedom of
defining only two processes among $\lossnorm_t$, $\defratenorm_t$
and $\rec_t$. The more natural approach would be modeling
explicitly $(\defratenorm_t, R_t)$, obtaining $\lossnorm_t$, or
modeling explicitly $(\lossnorm_t, R_t)$, obtaining
$\defratenorm_t$, all of them adapted.

However, if we choose to model both $\lossnorm_t$ and
$\defratenorm_t$ as ${\cal G}_t$-adapted processes and to infer
$\rec_t$, we have to ensure that the resulting process $\rec_t$
implicit in (\ref{eq:afc:bis}) is indeed left-continuous (and hence ${\cal G}_t$-predictable).

Indeed, in some formulations the predictability of the recovery is
not possible. It is also a notion not always realistic: whether
one or 125 names default in instant $(t-dt,t]$ (i.e. $d C_t = 1$ or $d C_t = 125$, respectively), we
would be imposing the recovery $R_t$ to be the same in both cases and, in particular, to depend only on the information up to $t^-$.

However, under adapted-ness and left-continuity the recovery rate can be expressed also in terms of the intensities of the loss and default rate
processes. From equation
(\ref{eq:afc:bis}), by definition of compensator, we obtain
\begin{equation}
\label{eq:recovery} \rec_t =
1-\frac{h_\lossnorm(t)}{h_\defratenorm(t)}.
\end{equation}
Equation (\ref{eq:recovery}) shows that the recovery rate at
default is directly related to the intensities of both the loss
and the default rate processes. Thus, the choice for the intensity
dynamics does induce a dynamics for the recovery rate.

In Brigo, Pallavicini and Torresetti (2006b) the cumulated
portfolio loss process $\lossnorm$ is directly modelled as a
GPL-type process with deterministic intensities and an extended
set of allowed jump amplitudes that go beyond $1/M,2/M,\ldots,1$,
according to~(\ref{eq:gpl:cappingl}). In this approach the
recovery is implicitly defined. Numerical results show that
calibrations are better with respect to the choice of modeling the
default counting process as a GPL process instead (Brigo,
Pallavicini, and Torresetti (2006a)). The direct loss modeling allows
for both portfolio total loss and for more granular small-size
losses. In particular super-senior tranches seem to be quoted
taking into account the possibility of portfolio total loss, so
that the direct loss model outperforms the default counting
process model with a constant or simple recovery formulation.

On the other hand, the GPCL model derived within the CPS framework
in the preceding sections requires direct modeling of the pool
counting process. Thus, if the recovery rate $R$ is constant, the
portfolio total loss is forbidden, since bounded to be not greater
than $1-R$ on a unit portfolio notional.

We now examine possible ways to model the loss more realistically,
starting from a GPL or GPCL model formulated in terms of default
counting process. This amounts to implicitly model the recovery
rate, since the number of defaults and the loss are linked by the
recovery at default.

\subsection{Recovery dynamics through Deterministic mapping}

A first approach to implicitly model recovery rates consists in
defining the cumulated portfolio loss $\lossnorm_t$ process as a
deterministic function of the pool counting process
$\defratenorm_t$ via a deterministic map, as previously done when
dealing with repeated defaults exceeding the pool size, through
the a-posteriori capping technique used in the basic GPL model
(see Section~\ref{sec:firstgpl}). Generalizing that approach leads
to the setting  \[\lossnorm_t:=\psi(\defratenorm_t), \] where
$\psi$ is a non-decreasing deterministic function with $\psi(0)=0$
and $\psi(1) \le 1$. What does this imply in terms of recovery
dynamics? We can easily write

\[   d \lossnorm_t = \sum_{k=1}^M \left[ \frac{\psi(\defratenorm_{t^-} +
k/M)-\psi(\defratenorm_{t^-})}{k/M}\right] 1_{\{ d \defratenorm_{t} =
k/M\}} d
\defratenorm_t \]
which shows that the recovery at default in this case would not be
predictable, depending explicitly from $dC_t$, except
for very special $\psi$'s.

A generalization based on a random process transformation (rather
than a deterministic function) of the counting process leading to
an implicit dynamics of the recovery process is presented in the
next section.

\subsection{Recovery dynamics through Gamma mapping}

Consider a stochastic process $u \mapsto \Psi_u$ in time $u$,
${\cal G}_u$-adapted and taking values in $[0,1]$, right-continuous with left limit, and independent of
the default counting process $\defratenorm_t$, and use it to map
the positive non-decreasing pool counting process $\defratenorm_t$
taking values in $[0,1]$ into the portfolio cumulated loss
$\lossnorm_t$, sharing the same characteristics, i.e. define
\[\lossnorm_t:=\Psi_{\defratenorm_t} . \]

Further, assume the process satisfies the following requirements,
enforcing the no-arbitrage conditions:
\[
\Psi_{0} = 0,\quad \Psi_{1} \le 1,\quad {\rm and} \quad d
 \Psi_t \ge 0
\]
This way the cumulated portfolio loss can be viewed as a
stochastic time change of the process $\Psi$. Further, in order to
allow for portfolio total loss, we enforce the stronger condition
\[
\Psi_{1} = 1 .
\]

The time change does not spoil the analytical tractability of the
model. If we know the probability distribution function of the
pool counting process and of $\Psi$, we can simply derive the
probability distribution function of the portfolio loss through an
iterated expectation, thanks to independence:
\[
\Qx{\lossnorm_t \le x} = \Ex{}{ \Qx{\lossnorm_t \le x |
\defratenorm_t}} = \int \Qx{\Psi_y \le x} p_{\defratenorm_t}(y)  dy
\]
%

As a relevant example, assume the process $u \mapsto \Psi_u$ is a
Gamma process with shape parameter $\mu(u)$ and scale parameter
$\nu$. The monotonicity of the resulting loss process can be
easily checked, while the probability distribution of the process
can be calculated explicitly. Indeed, as a direct calculation can
show, for any times $s<t<T$, the conditional distribution of
$\Psi_t$, given $\Psi_s$ and $\Psi_T$ is known in terms of the
Beta distribution.

The calculation of the unconditional distribution of the cumulated
portfolio loss follows directly.

Exactly as for the previous case based on the deterministic
transform $\psi$, here the implicit recovery at default turns out
to be not predictable in general.

\section{Conclusions}
We have extend the common Poisson shock (CPS) framework in two possible ways  that avoid repeated defaults. The second way, more consistent with the original spirit of the CPS framework, leads to the Generalized-Poisson adjusted-Cluster-dynamics Loss model (GPCL) . We have illustrated the relationship of the GPCL with our earlier Generalized Poisson Loss (GPL) model, pointing out that while the GPCL model shares the good calibration power of the GPL model, it further allows for consistency with single names, thus constituing one of the few explict examples of top down approaches to loss modeling with real consistency for single names, or of bottom up approaches with real dynamical features.

Further research concerns recovery dynamics, calibration and analysis of forward start tranches and tranche options, when liquid quotes will be available, and analysis of calibration stability through history. A preliminary analysis of stability with the GPL model is however presented in Brigo, Pallavicini and Torresetti (2006b), showing good results. This is encouraging and leads to assuming the GPCL stability as well, although a rigorous check is in order in further work.

\newpage

\appendix

\section{Market quotes\label{sec:market}}

The most liquid multi-name credit instruments available in the
market are credit indices and CDO tranches (e.g. DJi-TRAXX, CDX).

\subsection{Credit indices}

The index is given by a pool of names $1,2,\ldots,M$,
typically $M=125$, each with notional $1/M$ so that the total pool
has unitary notional. The index default leg consists of protection
payments corresponding to the defaulted names of the pool. Each
time one or more names default the corresponding loss increment is
paid to the protection buyer, until final maturity $T=T_b$ arrives
or until all the names in the pool have defaulted.

In exchange for loss increase payments, a periodic premium with
rate $\cds$ is paid from the protection buyer to the protection
seller, until final maturity $T_b$. This premium is computed on a
notional that decreases each time a name in the pool defaults, and
decreases of an amount corresponding to the notional of that name
(without taking out the recovery).

We denote with $\lossnorm_t$ the portfolio cumulated loss and
with $\defratenorm_t$ the number of defaulted names up to time
$t$ re-scaled by $M$. Thus, $0 \le\lossnorm_t\le\defratenorm_t\le 1$.
The discounted payoff of the two legs of the index is given as follows:
\[
\defleg(0) := \int_0^T D(0,t) d\lossnorm_t
\]
\[
\prmleg(0) :=  \cds_0 \sum_{i=1}^b D(0,T_i) \int_{T_{i-i}}^{T_i} (1-\defratenorm_t )dt
\]
where $D(s,t)$ is the (deterministic) discount factor between
times $s$ and $t$. The integral on the right hand side of the
premium leg is the outstanding notional on which the premium is
computed for the index. Often the premium leg integral involved in
the outstanding notional is approximated so as to obtain
\begin{eqnarray*}
\prmleg(0) = \cds_0 \sum_{i=1}^b \,\delta_i D(0,T_i) (1-\defratenorm_{T_i})
\end{eqnarray*}
where $\delta_i=T_i-T_{i-1}$ is the year fraction.

Notice that, differently from what will happen with the tranches
(see the following section), here the recovery is not considered
when computing the outstanding notional, in that only the number
of defaults matters.

The market quotes the value of $\cds_0$ that,
for different maturities, balances the two legs. If one has a
model for the loss and the number of defaults one may impose that
the loss and number of defaults in the model, when plugged inside
the two legs, lead to the same risk neutral expectation (and thus
price) when the quoted $\cds_0$ is inside the
premium leg, leading to

\begin{eqnarray}
\label{eq:index}
\cds_0 = \frac{\Ex{0}{\int_0^T D(0,t) d\lossnorm_t}}
        {\Ex{0}{\sum_{i=1}^b \,\delta_i D(0,T_i) (1-\defratenorm_{T_i})}}
\end{eqnarray}

\subsection{CDO tranches}

Synthetic CDO with maturity $T$ are contracts involving a
protection buyer, a protection seller and an underlying pool of
names. They are obtained by putting together a collection of
Credit Default Swaps (CDS) with the same maturity on different
names, $1,2,...,M$, typically $M=125$, each with notional $1/M$,
and then ``tranching" the loss of the resulting pool between
the points $A$ and $B$, with $0 \le A<B \le 1$.
\[
\losstr_t:= \frac{1}{B-A}\left[(\lossnorm_t-A) \ind{A<\lossnorm_t \le B}+(B-A)\ind{\lossnorm_t>B}\right]
\]

Once enough names have defaulted and the loss has reached $A$, the
count starts. Each time the loss increases the corresponding loss
change re-scaled by the tranche thickness $B-A$  is paid to the
protection buyer, until maturity arrives or until the total pool
loss exceeds $B$, in which case the payments stop.

The discounted default leg payoff can then be written as
\[
\defleg(0;A,B) := \int_0^T D(0,t) d\losstr_t
\]
Again, one should not  be confused by the integral, the loss
$\losstr_t$ changes with discrete jumps. Analogously,
also the total loss $\lossnorm_t$ and the tranche outstanding
notional change with discrete jumps.

As usual, in exchange for the protection payments, a premium rate
$\cdstr_0$, fixed at time $T_0=0$, is paid periodically, say at
times $T_1,T_2,\ldots,T_b=T$. Part of the premium can be paid at
time $T_0=0$ as an upfront $\upfronttr_0$.
The rate is paid on the
``survived" average tranche notional.
If we assume that the payments are made on the notional remaining
at each payment date $T_i$, rather than on the average in $[T_{i-1},T_i]$,
the discounted payoff of the premium leg can be written as
\begin{equation*}
\prmleg(0;A,B) := \upfronttr_0 +
                    \cdstr_0 \sum_{i=1}^b \, \delta_i D(0,T_i) (1-\losstr_{T_i})
\end{equation*}
where $\delta_i=T_i-T_{i-1}$ is the year fraction.

When pricing CDO tranches, one is interested in the premium rate
$\cdstr_0$ that sets to zero the risk neutral price of the
tranche. The tranche value is computed taking the (risk-neutral)
expectation (in $t=0$) of the discounted payoff consisting on the
difference between the default and premium legs above. We obtain
\begin{eqnarray}
\label{eq:tranche}
\cdstr_0 = \frac{\Ex{0}{\int_0^T D(0,t) d\losstr_t} - \upfronttr_0}
                {\Ex{0}{\sum_{i=1}^b \, \delta_i D(0,T_i) (1-\losstr_ {T_i})}}
\end{eqnarray}
The above expression can be easily recast in terms of the upfront
premium $\upfronttr_0$ for tranches that are quoted in terms of upfront fees.

The tranches that are quoted on the market refer to standardized
pools, standardized attachment-detachment points $A-B$ and
standardized maturities $T$.

Actually, for the i-Traxx and CDX pools, the equity tranche $(A=0, B=3\%)$ is quoted by means of the fair $\upfronttr_0$, while assuming $\cdstr_0 = 500 bps$.
All other tranches are quoted by means of the fair $\cdstr_0$, assuming no upfront fee  ($\upfronttr_0=0)$.


\section{Tables: Calibration Inputs and Outputs}

\begin{table}[h]
\begin{center}\small
\begin{tabular}{|cc|cc|cc|cc|}\hline
{\bf Date} & {\bf Rate} & {\bf Date} & {\bf Rate} & {\bf Date} & {\bf Rate} & {\bf Date} & {\bf Rate} \\\hline
 20-Dec-06 &     3.41\% &  21-Sep-09 &     3.71\% &  20-Jun-12 &     3.75\% &  20-Mar-15 &     3.83\% \\
 20-Mar-07 &     3.57\% &  21-Dec-09 &     3.72\% &  20-Sep-12 &     3.76\% &  22-Jun-15 &     3.84\% \\
 20-Jun-07 &     3.66\% &  22-Mar-10 &     3.72\% &  20-Dec-12 &     3.76\% &  21-Sep-15 &     3.84\% \\
 20-Sep-07 &     3.70\% &  21-Jun-10 &     3.72\% &  20-Mar-13 &     3.77\% &  21-Dec-15 &     3.85\% \\
 20-Dec-07 &     3.72\% &  20-Sep-10 &     3.72\% &  20-Jun-13 &     3.77\% &  21-Mar-16 &     3.86\% \\
 20-Mar-08 &     3.72\% &  20-Dec-10 &     3.72\% &  20-Sep-13 &     3.78\% &  20-Jun-16 &     3.87\% \\
 20-Jun-08 &     3.72\% &  21-Mar-11 &     3.73\% &  20-Dec-13 &     3.79\% &  20-Sep-16 &     3.87\% \\
 22-Sep-08 &     3.72\% &  20-Jun-11 &     3.73\% &  20-Mar-14 &     3.80\% &  20-Dec-16 &     3.88\% \\
 22-Dec-08 &     3.72\% &  20-Sep-11 &     3.74\% &  20-Jun-14 &     3.80\% &            &            \\
 20-Mar-09 &     3.71\% &  20-Dec-11 &     3.74\% &  22-Sep-14 &     3.81\% &            &            \\
 22-Jun-09 &     3.71\% &  20-Mar-12 &     3.74\% &  22-Dec-14 &     3.82\% &            &            \\\hline
\end{tabular}
\caption{\label{input:discount} EUR zero-coupon
continuously-compounded spot rates (ACT/365).}
\end{center}
\end{table}

\newpage

\begin{table}[h]
\begin{center}\small
\begin{tabular}{|c|c|cccc|}\hline
              & {\bf Att-Det} &            \multicolumn{ 4}{|c|}{{\bf Maturities}} \\
              &            &        3y &          5y &          7y &         10y \\\hline
{\bf Index}   &            & ~~18(0.5) &  ~~30(0.5)  & ~~~~40(0.5) & ~~~~51(0.5) \\\hline
{\bf Tranche} &        0-3 & ~350(150) &   ~1975(25) & ~~3712(25)~ & ~~4975(25)~ \\
              &        3-6 & 5.50(4.0) & 75.00(1.0)  & 189.00(2.0) & 474.00(4.0) \\
              &        6-9 & 2.25(3.0) & 22.25(1.0)  & ~54.25(1.5) & 125.50(3.0) \\
              &       9-12 &           & 10.50(1.0)  & ~26.75(1.5) & ~56.50(2.0) \\
              &      12-22 &           & ~4.00(0.5)  & ~~9.00(1.0) & ~19.50(1.0) \\
              &     22-100 &           & ~1.50(0.5)  & ~~2.85(0.5) & ~~3.95(0.5) \\\hline
\end{tabular}
\caption{\label{input:tranche} DJi-TRAXX index and tranche quotes
in basis points on October 2, 2006, along with the bid-ask spreads.
Index and tranches are quoted through the
periodic premium, whereas the equity tranche is quoted as an
upfront premium. See Appendix \ref{sec:market}.}
\end{center}
\end{table}

\begin{table}
\begin{center}\small
\begin{tabular}{|c|c|cccc|}\hline
              & {\bf Att-Det} &            \multicolumn{ 4}{|c|}{{\bf Maturities}} \\
              &            &        3y &          5y &         7y &         10y \\\hline
{\bf Index}   &            & ~~24(0.5) & ~~~~40(0.5) & ~~~~49(0.5)~ & ~~~~61(0.5)~ \\\hline
{\bf Tranche} &        0-3 & ~975(200) & ~~3050(100) & ~~4563(200)~ & ~~5500(100)~ \\
              &        3-7 & 7.90(1.6) & 102.00(6.1) & 240.00(48.0) & 535.00(21.4) \\
              &       7-10 & 1.20(0.2) & ~22.50(1.4) & ~53.00(10.6) & 123.00(7.4)~ \\
              &      10-15 & 0.50(0.1) & ~10.25(0.6) & ~23.00(4.6)~ & ~59.00(3.5)~ \\
              &      15-30 & 0.20(0.1) & ~~5.00(0.3) & ~~7.20(1.4)~ & ~15.50(0.9)~ \\\hline
\end{tabular}
\caption{\label{input:tranchecdx} CDX index and tranche quotes
in basis points on October 2, 2006, along with the bid-ask spreads.
Index and tranches are quoted through the
periodic premium, whereas the equity tranche is quoted as an
upfront premium. See Appendix \ref{sec:market}.}
\end{center}
\end{table}

\begin{table}[h]
\begin{center}\small
\begin{tabular}{|c|cccc|}\hline &  &  &  &  \\
   {\boldmath$\alpha_j$} &     \multicolumn{ 4}{|c|}{{\boldmath${\Lambda}^0_j(T)$}} \\\hline
           &         3y &         5y &         7y &        10y \\
         1 &      0.778 &      1.318 &      3.320 &      4.261 \\
         3 &      0.128 &      0.536 &      0.581 &      1.566 \\
        15 &      0.000 &      0.004 &      0.024 &      0.024 \\
        19 &      0.000 &      0.007 &      0.011 &      0.028 \\
        32 &      0.000 &      0.000 &      0.000 &      0.007 \\
        79 &      0.000 &      0.000 &      0.003 &      0.003 \\
       120 &      0.000 &      0.002 &      0.003 &      0.008 \\\hline
\end{tabular}
\hspace*{1cm}
\begin{tabular}{|c|cccc|}\hline &  &  &  &  \\
   {\boldmath$\alpha_j$} &     \multicolumn{ 4}{|c|}{{\boldmath$ {M \choose \alpha_j} {\widetilde{\Lambda}}_j(T)$}} \\\hline
           &         3y &         5y &         7y &        10y \\
         1 &      0.882 &      1.234 &      3.223 &      3.661 \\
         3 &      0.128 &      0.615 &      0.682 &      1.963 \\
        15 &      0.001 &      0.002 &      0.023 &      0.023 \\
        19 &      0.000 &      0.009 &      0.016 &      0.043 \\
        57 &      0.000 &      0.000 &      0.002 &      0.007 \\
        80 &      0.000 &      0.000 &      0.000 &      0.010 \\
       125 &      0.001 &      0.005 &      0.042 &      0.042 \\\hline
\end{tabular}
\caption{\label{tab:param} DJi-TRAXX pool.  Left side: cumulated intensities,
integrated up to tranche maturities, of the basic GPL model. Each
row $j$ corresponds to a different Poisson component with jump
amplitude $\alpha_j$. Right side: cumulated
cluster intensities, integrated up to tranche maturities, and
multiplied by the number of clusters of the same size at time $0$.
Each row $j$ corresponds to a different cluster size $\alpha_j$.
The amplitudes/cluster-sizes not listed have an intensity
below $10^{-7}$. The recovery rate is $40\%$. }
\end{center}
\end{table}

\begin{table}[h]
\begin{center}\small
\begin{tabular}{|c|cccc|}\hline &  &  &  &  \\
   {\boldmath$\alpha_j$} &     \multicolumn{ 4}{|c|}{{\boldmath${\Lambda}^0_j(T)$}} \\ \hline
           &         3y &         5y &         7y &        10y \\
         1 &      1.132 &      3.043 &      4.247 &      7.166 \\
         2 &      0.189 &      0.189 &      0.812 &      1.625 \\
         6 &      0.011 &      0.091 &      0.091 &      0.091 \\
        18 &      0.000 &      0.006 &      0.028 &      0.028 \\
        23 &      0.000 &      0.004 &      0.005 &      0.032 \\
        32 &      0.000 &      0.000 &      0.000 &      0.009 \\
       124 &      0.000 &      0.003 &      0.005 &      0.010 \\\hline\end{tabular}
\hspace*{1cm}
\begin{tabular}{|c|cccc|}\hline   &  &  &  &  \\
   {\boldmath$\alpha_j$} &     \multicolumn{ 4}{|c|}{{\boldmath$ {M \choose  \alpha_j} \widetilde{\Lambda}_j(T)$}} \\ \hline
           &         3y &         5y &         7y &        10y \\
         1 &      0.063 &      0.552 &      3.100 &      6.661 \\
         2 &      0.804 &      1.531 &      1.531 &      2.076 \\
         3 &      0.020 &      0.195 &      0.195 &      0.195 \\
        17 &      0.000 &      0.010 &      0.037 &      0.087 \\
        32 &      0.000 &      0.003 &      0.009 &      0.032 \\
       110 &      0.000 &      0.000 &      0.000 &      0.010 \\
       125 &      0.000 &      0.011 &      0.054 &      0.054 \\\hline
\end{tabular}
\caption{\label{tab:paramcdx} CDX pool. Left side: cumulated intensities,
integrated up to tranche maturities, of the basic GPL model. Each
row $j$ corresponds to a different Poisson component with jump
amplitude $\alpha_j$. Right side: cumulated
cluster intensities, integrated up to tranche maturities, and
multiplied by the number of clusters of the same size at time $0$.
Each row $j$ corresponds to a different cluster size $\alpha_j$.
The amplitudes/cluster-sizes not listed have an intensity
below $10^{-7}$. The recovery rate is $40\%$. }
\end{center}
\end{table}

\begin{table}[h]
\begin{center}\small
\begin{tabular}{|c|c|cccccc|}\hline
           &            &                                        \multicolumn{ 6}{|c|}{{\bf Tranches}} \\
{\bf Models} & {\bf Maturities} &        0-3 &        3-6 &        6-9 &       9-12 &      12-22 &     22-100 \\\hline
 {\bf ITL} &         3y &     18.6\% &      0.2\% &      0.1\% &      0.0\% &      0.0\% &      0.0\% \\
    {\bf } &         5y &     44.6\% &      4.2\% &      1.2\% &      0.6\% &      0.2\% &      0.1\% \\
    {\bf } &         7y &     71.0\% &     14.5\% &      4.3\% &      2.1\% &      0.7\% &      0.2\% \\
    {\bf } &        10y &     91.6\% &     49.2\% &     14.1\% &      6.4\% &      2.2\% &      0.4\% \\\hline
 {\bf GPL} &         3y &     18.6\% &      0.2\% &      0.1\% &      0.1\% &      0.0\% &      0.0\% \\
     {\bf } &         5y &     44.5\% &      4.2\% &      1.2\% &      0.6\% &      0.2\% &      0.1\% \\
    {\bf } &         7y &     70.8\% &     14.6\% &      4.3\% &      2.1\% &      0.7\% &      0.2\% \\
    {\bf } &        10y &     91.2\% &     47.2\% &     14.6\% &      6.4\% &      2.2\% &      0.4\% \\\hline
{\bf GPCL} &         3y &     18.7\% &      0.2\% &      0.1\% &      0.0\% &      0.0\% &      0.0\% \\
    {\bf } &         5y &     44.7\% &      4.2\% &      1.2\% &      0.6\% &      0.2\% &      0.1\% \\
    {\bf } &         7y &     70.9\% &     14.6\% &      4.3\% &      2.1\% &      0.7\% &      0.2\% \\
    {\bf } &        10y &     91.2\% &     47.5\% &     14.5\% &      6.4\% &      2.2\% &      0.4\% \\\hline
\end{tabular}
\caption{\label{tab:itl} Implied expected tranched loss for the
ITL, GPL and GPCL models. Results refer to DJi-TRAXX market.}
\end{center}
\end{table}

\begin{table}[h]
\begin{center}\small
\begin{tabular}{|c|c|cc|}\hline
           & {\bf Att-Det} & \multicolumn{ 2}{|c|}{{\bf DJi-TRAXX 10y}} \\
           &            &  {\bf GPL} & {\bf GPCL} \\\hline
{\bf Index} &           &       0.00 &       0.00 \\\hline
{\bf Tranche} &     0-3 &       0.76 &       0.62 \\
           &        3-6 &      -2.35 &      -1.93 \\
           &        6-9 &       1.21 &       1.04 \\
           &       9-12 &      -0.40 &      -0.36 \\
           &      12-22 &       0.02 &       0.02 \\
           &     22-100 &       0.00 &       0.00 \\\hline
\end{tabular}
\hspace*{1cm}
\begin{tabular}{|c|c|cc|}\hline
           & {\bf Att-Det} & \multicolumn{ 2}{|c|}{{\bf CDX 10y}} \\
           &            &  {\bf GPL} & {\bf GPCL} \\\hline
{\bf Index} &           &       0.00 &      -0.06 \\\hline
{\bf Tranche} &     0-3 &       1.43 &       1.60 \\
           &        3-7 &      -0.45 &      -0.22 \\
           &       7-10 &       0.22 &       0.25 \\
           &      10-15 &      -0.08 &      -0.12 \\
           &      15-30 &       0.01 &       0.07 \\
           &            &            &            \\\hline
\end{tabular}
\caption{\label{tab:cal} Calibration errors calculated with
the GPL and GPCL models with respect to the bid-ask spread
(i.e. $\epsilon_i$ in (\ref{eq:calibrerror})) for tranches
quoted by the market for the ten year maturity
(see Tables \ref{input:tranche} and \ref{input:tranchecdx}).
The left panel refers to DJi-TRAXX market quotes, while the right
panel refers to CDX market quotes.
Calibration errors for the other maturities are within the
bid-ask spread and therefore they are not reported.
The recovery rate is $40\%$ .}

\end{center}
\end{table}

\end{document}